\documentclass[aps,prc,twocolumn,groupedaddress]{revtex4-1}
\usepackage{bm}
\usepackage{color}
\usepackage{graphicx}
\usepackage{amssymb,amsmath}
\allowdisplaybreaks
\newcommand{\EDM}{\hat{\boldsymbol{D}}}
\newcommand{\sone}{\boldsymbol{\sigma}_1}
\newcommand{\stwo}{\boldsymbol{\sigma}_2}
\newcommand{\stre}{\boldsymbol{\sigma}_3}
\newcommand{\tone}{\boldsymbol{\tau}_1}
\newcommand{\ttwo}{\boldsymbol{\tau}_2}
\newcommand{\ttre}{\boldsymbol{\tau}_3}

\newcommand{\vtrv}{V_{TRV}}
\def\a{\alpha}

\def\omp{\omega_+}
\def\omm{\omega_-}
\def\bmr{{\bm r}}
\def\bmvr{\hat{\bm r}}
\def\bmx{{\bm x}}
\def\bmp{{\bm p}}
\def\bmk{{\bm k}}
\def\bmq{{\bm q}}
\def\bmQ{{\bm Q}}
\def\bmR{{\bm R}}
\def\bmK{{\bm K}}

\def\bmP{{\bm P}}
\def\bmL{{\bm L}}
\def\fp{f_\pi}
\def\ga{g_A}
\def\mp{m_\pi}
\def\deltatre{\Delta_3}
\def\omk{\omega_k}
\def\det{{{}^2{\rm H}}}
\def\tri{{{}^3{\rm H}}}
\def\hel{{{}^3{\rm He}}}
\def\Lx{\Lambda_{\chi}}
\newcommand{\bmsi}{{\bm \sigma}}
\newcommand{\bmna}{{\bm \nabla}}
\newcommand{\lchi}{\Lambda_\chi}
\newcommand{\bra}{\langle}
\newcommand{\ket}{\rangle}
\def\TO{T_0}
\def\ttp{T_1^+}
\def\ttm{T_1^-}
\def\tten{T_2}
\def\deps{d_\epsilon}
\def\ssp{\hat{S}^+_\bmr}
\def\ssm{\hat{S}^-_\bmr}
\def\slp{\hat{S}^+_{L}}
\def\slm{\hat{S}^-_{L}}
\def\srx{\hat{S}^\times_\bmr}
\def\srlp{\hat{S}^+_{\bmr L}}
\def\srlm{\hat{S}^-_{\bmr L}}
\def\m{\phantom{-}}

\begin{document}


\title{Time Reversal Violation in Light Nuclei}


\author{A.\ Gnech$^{\,{\rm a,b}}$, M.\ Viviani$^{\,{\rm b}}$}
\affiliation{
$^{\rm a}$\mbox{Gran Sasso Science Institute, 67100 L'Aquila, Italy}\\
$^{\rm b}$\mbox{INFN-Pisa, 56127 Pisa, Italy}
}


\date{\today}

\begin{abstract}
  Time Reversal Violation (TRV) interactions between quarks which appear in
  Standard Model (SM) and beyond-SM theories can induce TRV components
  in the nucleon-nucleon potential. The effects of these components can be
  studied by measuring the electric dipole moment (EDM) of light nuclei.
  In this work we present a complete derivation of the TRV nucleon-nucleon
  and three-nucleon potential up to next-to-next-to leading order (N2LO) in
  a chiral effective field theory ($\chi$EFT) framework. The TRV interaction
  is then used to evaluate the EDM of $\det$, $\tri$ and $\hel$ focusing in
  particular on the effects of the TRV three-body force and on the calculation
  of the theoretical errors. In case of a measurement of the EDM of these
  nuclei, the result of present work would permit to determine the values of the low
  energy constants and to identify the source of TRV.
\end{abstract}

\pacs{}

\maketitle
\section{Introduction}

The violation of parity (P) with the conservation of charge conjugation
(C) generates a CP violation that, using the CPT theorem, results in
a time reversal violation (TRV). TRV is a key ingredient in the
explanation of the observed matter-antimatter asymmetry in the Universe
~\cite{AS67}.
The Standard Model (SM) has a natural source of CP-violation
in the Cabibbo-Kobayashi-Maskawa
(CKM) quark mixing matrix, however this mechanism is not sufficient for
explaining the observed asymmetry~\cite{AC93}.
This discrepancy opens a window in possible TRV effect
in the SM, such as the $\theta$-term in the Quantum
Chromodynamics (QCD) sector~\cite{GH76}, or in other sources of
beyond-SM (BSM) theories~\cite{MP05}.

The measurement of Electric Dipole Moments (EDMs) of particles is the most
promising observable for studying TRV beyond CKM mixing matrix effects. Indeed,
the EDM induced by the complex phase of the CKM matrix are suppressed since the EDM
does not involve flavour changing~\cite{MP05,AC97,MU12,MU13}.
Therefore, any non-vanishing EDM of a nuclear or an atomic
system would highlight  TRV effects beyond the CKM mixing matrix.
The present experimental upper bounds on the EDMs of neutron and proton
are $|d_n|<2.9\cdot 10^{-13}\ e$ fm~\cite{CA06}
and $|d_p|<7.9\cdot 10^{-12}\ e$ fm,
where the proton EDM has been inferred from a measurement of the diamagnetic
${}^{199}{\rm Hg}$ atom~\cite{WC09} using a calculation of the nuclear
Schiff moment~\cite{VF03}.
For the electron, the most recent upperbound is
$|d_e|<8.7\cdot10^{-16}\ e$ fm~\cite{JB14},
derived from the EDM of the ThO molecule.

In this context, there are proposals for the direct measurement of EDMs of
electrons, single nucleons and light nuclei in dedicated storage rings
~\cite{YF06,YK11,AL13,JP13,FR13}. This new
approach plans to reach an accuracy of $\sim 10^{-16}\ e$ fm,  improving
the sensitivity in particular in the hadronic sector. Any measurement of a
non-vanishing EDM of this magnitude would be the evidence of
TRV beyond CKM effects. However,
a single measurement will not be sufficient to identify the source of TRV.
For this reason, the measurement of EDM of various light nuclei such as $\det$,
$\tri$ and $\hel$ can impose constrains on the TRV sources.

On the other hand, also spin rotation of
polarized neutrons along the $y$-axis
can be used as probe for TRV~\cite{PK82,LS82,CP04,CP06,JDB14}.
This explorative study is motivated by the new
sensitivity reached by various cold neutron facilities such as
the Los Alamos
Neutron Science Center, the National Institute of Standards
and Technology (NIST) Center for Neutron Research, and
the Spallation Neutron Source (SNS) at Oak Ridge National
Laboratory.
Even if the sensitivity of the spin-rotation on TRV is far away from the present
sensitivity of  the EDM, this observable addresses the same physics of the EDM,
posing, for the future, as a complementary and independent test for
TRV in nuclei. Also other TRV observables in proton- and
neutron-nucleus scattering have been proposed as probes of TRV effects~\cite{YH11,JDB14,YNU16,VG18,PF19}.

The use of light nuclei to study TRV results to be advantageous because
the nuclear physics of the systems is theoretically under control and so
the TRV effects can be easily highlighted. In particular, the chiral effective
field theory ($\chi$EFT) has provided a practical and successful scheme to study
two and many-nucleon interactions~\cite{DR15,EE15}
and the interaction of electroweak probes with nuclei
~\cite{Park96,Park03,Koelling09,Pastore09,Koelling11,Pastore11,Piarulli13}.
The $\chi$EFT approach is based on the observation that the chiral symmetry
exhibited by QCD has a noticeable impact in the low-energy regime. Therefore,
the form of the strong interactions of pions among themselves and other particles is
severely constrained by the transformation properties
of the fundamental Lagrangian
under Lorentz, parity, time-reversal, and chiral symmetry~\cite{W66,CCWZ69}.
The Lagrangian terms
can be organized as an expansion in powers of $Q/\Lx$,
where $\Lx\simeq1$ GeV specifies
the chiral symmetry breaking scale and $Q$ is the exchanged pion momentum.
Each term is associated to a low-energy
constants (LECs) which are then determined by fits to experimental data.

The $\chi$EFT method permits to construct also an effective TRV Lagrangian
treating all possible 
sources of TRV. The TRV Lagrangian induced by the $\theta$-term was derived
in Refs.~\cite{EM10,JB150}. Also BSM terms such as supersymmetry, multi-Higgs
scenarios, left-right symmetric model {\it etc.} induce TRV operators at the
quark-gluon level which appear at level of dimension six (see for
example Ref.~\cite{BG10}). The $\chi$EFT
Lagrangians for these sources were derived in Refs.~\cite{JB150,JV13,WD13,JV11}.
This approach permits not only
to determine the TRV interactions but also to estimate the chiral order of the
LECs and their values as function of the fundamental parameters, providing a
direct connection between the fundamental theories and the nuclear observables
~\cite{EM10,JB150,JV13}.
To be noticed that the chiral order of the Lagrangian terms, which is determined by
the products of the chiral order of the dynamical part and that of the LECs, really
depends on the particular source of TRV.
Therefore, when the LECs will be determined experimentally it will be possible
to identify the dynamical properties of the TRV source~\cite{JV11,WD14}.

Starting from the TRV Lagrangian, de Vries {\it et al.}~\cite{CM11} and also Bsaisou
{\it et al.}~\cite{JB13}
derived the chiral potential up to next-to-next leading order (N2LO)
including only nucleon-pion interaction and
contact interactions. In both works also the electromagnetic currents which
play a role at N2LO for the EDM were derived but only in Ref.~\cite{JB13}
they were used
to evaluate the EDM of the deuteron. In Ref.~\cite{JV11b}
the calculation of the EDM
of $\det$, $\tri$ and $\hel$
was performed using only the one-pion-exchange part of the TRV potential coupled 
with phenomenological potential for the parity conserving (PC) part of the interaction.

Subsequent works showed the presence in the TRV Lagrangian of a three-pion
term~\cite{JB150}, which was included in the calculation
for the first time by Bsaisou {\it et al.}~\cite{JB151}.
This term generates at next-to-leading order (NLO) also a TRV three body
force, which contribution to the  $\tri$ and $\hel$ EDM
was found to be smaller than expected by the chiral counting.
The calculation reported in Ref.~\cite{JB151} was also the first to use a complete
chiral approach including the TRV potential up to NLO  and
the PC potential up to N2LO.

The aim of this work is twofold:
First, the construction of a TRV potential up to N2LO
considering all possible TRV interaction terms in the $\chi$EFT
without making any assumption for the chiral order of the LECs.
In this way all possible
sources of TRV can be studied just setting the LECs to their estimated values and
turning on and off the various terms in the Lagrangian.
The second is the study of the EDM of $\det$, $\tri$ and $\hel$,
providing a suitable framework for the future determination of the
LECs. In our calculation the contribution of the TRV three-body force
is found to be sizably larger than reported in Ref.~\cite{JB151}. 

Finally, it is worthy to mention that exists a different approach to the
derivation of the TRV
nuclear forces based on meson-exchange model~\cite{CP04}.
This model includes pion and
vector-meson exchange with 10 unknown meson constants. Such a theory,
which has a wider energy range of validity but it is less systematic and
with no direct connection with the fundamental Lagrangian,
has been used to study the EDM of
light nuclei~\cite{CP04,IS08,NY15,YH13} and the neutron
spin rotation $\vec{n}-\vec{p}$~\cite{CP06} and $\vec{n}-\vec{d}$
~\cite{CP04,YH11} scattering.

The present paper is organized as follows. In Sec.~\ref{sec:trvlag}
we will present the TRV chiral Lagrangian up to order $Q^2$
relevant for the calculation of the TRV potential, while
in Sec.~\ref{sec:trvpot} we derive the TRV potential at N2LO. In
Sec.~\ref{sec:res}, we report the results obtained for the
EDM of $\det$, $\tri$ and $\hel$ using the N2LO potential.
Finally, in Sec.~\ref{sec:conc} we present our conclusions and
perspectives. The technical details relative to the contributions of the various
 diagrams, and of the derivation of the
potential in configuration space are given in Appendices~\ref{app:vertex},
~\ref{app:pot} and~\ref{app:rpot}. Moreover, in Appendix~\ref{app:NNNconv} we
will give some details of the calculation of the
trinucleon wave function negative-parity component and about the
convergence of the TRV three body force contribution.

\section{The TRV Lagrangian}\label{sec:trvlag}

The various possible sources of TRV in SM induce a TRV $NN$ and $NNN$ potential.
This potential can be constructed starting from a pion-nucleon effective
Lagrangian which includes, in principle, an infinite set of terms
which violates the chiral symmetry as the fundamental (quark-level)
Lagrangian. The effective Lagrangian can be ordered by a power
counting scheme which permits to select the most important interactions.
In literature the chiral order of the TRV Lagrangian is determined by considering
the estimated order of the LECs~\cite{EM10,JV13,JB150} which, however,
is source dependent.

In this section we present only the TRV Lagrangian terms which can give some
contribution to TRV $NN$ and $NNN$ potential up to N2LO in terms of pion field.
In order to remain source independent we consider
isoscalar, isovector and isotensor
terms and we determine the chiral order of the TRV Lagrangian
considering only the dynamical part.
Namely, in the following we will consider all
the TRV LECs to be equally important. To deal with a
specific source of TRV, it will be sufficient to set
some of the LECs to be zero, etc.
At order $Q^0$ the TRV pion-nucleon Lagrangian includes three terms
~\cite{EM10,JB150}
\begin{eqnarray}
  {\cal{L}}_{\rm TRV}^{\pi N\,(0)}=g_0\overline{\psi}\vec{\pi}\cdot\vec{\tau}
  \psi+g_1\overline{\psi}\pi_3\psi+g_2\overline{\psi}\pi_3\tau_3\psi\,,
\end{eqnarray}
where $\vec{\pi}$ is the pion field and $\psi$ is the nucleon field. To be noticed
that the isotensor term is usually considered of higher order.
At the same order a purely pionic interaction appears~\cite{JB150}, which reads
\begin{eqnarray}
  {\cal{L}}_{\rm TRV}^{3\pi\,(0)}=M\Delta_3 \pi_3\pi^2\,,
\end{eqnarray}
where $M=938.88$ MeV is the average nucleon mass.
At order $Q$ we have only one term that can give contribution
to the potential up to N2LO and it reads, given explicitly as,
\begin{eqnarray}
  {\cal{L}}_{\rm TRV}^{\pi N\,(1)}=\frac{g_V^{(1)}}{2M\fp}\big[\overline{\psi}
    \partial_\mu(\vec{\pi}\times\vec{\tau})_3\partial^\mu\psi+\ h.c.\ \big]\,,
  \label{eq:ltrv1}
\end{eqnarray}
while no Lagrangian terms for the pure pionic interaction are allowed.
At order $Q^2$ we get six new contributions,
\bgroup
\arraycolsep=1.0pt
\begin{eqnarray}
  {\cal{L}}_{\rm TRV}^{\pi N\,(2)}&=&
   {g_{S1}^{(2)}\over\fp^2}\overline{\psi}\partial_\mu\partial^\mu
  (\pi\cdot\tau)\psi\nonumber\\
  &+&{g_{S2}^{(2)}\over 2M^2\fp^2}\big[\overline{\psi}\partial_\mu\partial_\nu
  (\pi\cdot\tau)\partial^\mu\partial^\nu\psi+{\rm h.c.} \big]\nonumber\\
  &+&{g_{V1}^{(2)}\over\fp^2}\overline{\psi}\partial_\mu\partial^\mu
  \pi_3\psi\nonumber\\
  &+&{g_{V2}^{(2)}\over 2M^2\fp^2}\big[\overline{\psi}\partial_\mu\partial_\nu
  \pi_3\partial^\mu\partial^\nu\psi+{\rm h.c.} \big]\nonumber\\
  &+&{g_{T1}^{(2)}\over\fp^2}\overline{\psi}\partial_\mu\partial^\mu
  \pi_3\tau_3\psi\nonumber\\
  &+&{g_{T2}^{(2)}\over 2M^2\fp^2}\big[\overline{\psi}\partial_\mu\partial_\nu
    \pi_3\tau_3\partial^\mu\partial^\nu\psi+{\rm h.c.} \big]\ ,
  \label{eq:ltrv2}
\end{eqnarray}
\egroup
where $S$, $V$, $T$ stand for isoscalar, isovector and isotensor 
respectively and $\fp\simeq92$ MeV is the pion decay constant.
The three-pion interaction that appears at this level gives contributions
of order $Q^2$ to the TRV nuclear potential which is beyond our purpose, therefore
we do not consider it.
The Lagrangian contains also four-nucleon contact terms
${\cal L}^{\rm CT}_{\rm TRV}$
representing interactions originating from excitation of 
resonances and exchange of heavy mesons. At lowest order
${\cal L}^{\rm CT}_{\rm TRV}$ contains only five independent
four-nucleon contact
terms with a single gradient.

In the following we also need the PC Lagrangian up to N2LO:
\bgroup
\arraycolsep=1.0pt
\begin{eqnarray}
  {\cal L}^{PC}&=& {\cal L}_{\pi\pi}^{(2)}+\ldots\nonumber\\
   &+&  {\cal L}_{N\pi}^{(1)}+{\cal L}_{N\pi}^{(2)}+{\cal
    L}_{N\pi}^{(3)}+\ldots+ {\cal L}^{PC}_{CT}\ ,\label{eq:Lpc} \\
  {\cal L}_{\pi\pi}^{(2)}&=& {f_\pi^2\over 4} \langle\nabla_\mu U^\dag
  \nabla^\mu U + \chi^\dag U+\chi U^\dag\rangle\ , \label{eq:Lpc_pipi2}\\
  {\cal L}_{N\pi}^{(1)} &=& \overline{\psi}\Bigl(i\gamma^\mu D_\mu -M
  +{g_A\over 2} \gamma^\mu\gamma^5 u_\mu \Bigr)\psi\
  , \label{eq:Lpc_pin1}\\
   {\cal L}_{N\pi}^{(2)}  &=& c_1 \overline{\psi}\langle \chi_+\rangle
   \psi  - \frac{c_2}{8M^2}\big[\overline{\psi}\langle u_\mu u_\nu \rangle
     D^\mu D^\nu \psi +\  {\rm h.c.}\big]\nonumber\\
   &+&c_3\overline{\psi}\frac{1}{2}\langle u_\mu u^\mu \rangle \psi+
   c_4 \overline{\psi}\frac{i}{4}[u_\mu,u_\nu]\sigma^{\mu\nu}\psi+
   \ldots \ ,\label{eq:Lpc_pin2}\\
  {\cal L}_{N\pi}^{(3)}  &=& d_{16} \overline{\psi}{1\over 2}
  \gamma^\mu\gamma^5 u_\mu \langle \chi_+\rangle \psi\nonumber \\
  &+&  d_{18} \overline{\psi}{i\over 2}
  \gamma^\mu\gamma^5 [D_\mu, \chi_-] \psi+\ldots\ ,\label{eq:Lpc_pin3}
\end{eqnarray}
\egroup
where the building blocks $U$, $u_\mu$, $\chi_+$, $\chi_-$ and the covariant
derivative $D_\mu$ are defined in Ref.~\cite{viviani14}.
We have omitted all the terms not
relevant in the present work (for the complete expression for Lagrangian
$ {\cal L}_{N\pi}^{(2)}$ and ${\cal L}_{N\pi}^{(3)}$ in Ref.~\cite{NF00}).
Four-nucleon contact terms (see,
for example, Refs.\cite{EE09,RM11}) are lumped into ${\cal L}^{PC}_{CT}$.
The parameters $c_1$, $c_2$, $c_3$, $c_4$, $d_{16}$, and $d_{18}$ are
LECs entering the PC Lagrangian. All the constants entering the terms
discussed in this section have to be considered as ``bare'' parameters
(i.e. unrenormalized).

\section{The TRV potential up to order Q}\label{sec:trvpot}
In this section, we discuss the derivation of the TRV $NN$ and $NNN$ potential
at N2LO. We provide, order by order in  power counting the formal expressions
for it in terms of the time-ordered perturbation theory (TOPT) amplitudes,
following the scheme presented in Ref.~\cite{viviani14}.
Then, the various diagrams associated with these amplitudes are discussed
(additional details are given in Appendix~\ref{app:pot}).
We will give also some hints in the renormalization of the coupling constants.

\subsection{From amplitudes to potentials}
We start considering the conventional $NN$ scattering amplitude
$\bra N'N' |T|NN\ket$, where $|NN\ket$ and $|N'N'\ket$ represent the initial
and final two nucleon state and $T$ can be written as,
\begin{eqnarray}
 T =  H_I \sum_{n=1}^\infty \left( 
 \frac{1}{E_i -H_0 +i\, \eta } H_I \right)^{n-1} \ ,
\label{eq:pt}
\end{eqnarray}
where $E_i$ is the initial energy of the two nucleons, $H_0$ is the Hamiltonian
describing free pions and nucleons, and $H_I$ is the Hamiltonian
describing interactions among these particles. To be noticed that
in Eq.~(\ref{eq:pt}) 
the interaction Hamiltonian $H_I$ is in the Schr\"odinger picture and 
that, at the order of interest here, it follows
simply from $H_I=-\int {\rm d} \bmx\; {\cal L}_I (t=0,\bmx)$,
where ${\cal L}_I$ is the interaction Lagrangian in interaction
picture.  Vertices from $H_I$
are listed in Appendix~\ref{app:vertex}.

The $NN$ scattering amplitude can be organized as
an expansion in powers of $Q/\lchi\ll 1$, where $\lchi\simeq1$ GeV is the typical
hadronic mass scale,
\begin{equation}
  \bra N'N' |T|NN\ket=\sum_n T^{(n)}\,,
  \label{eq:teft}
\end{equation}
where $T^{(n)}\sim Q^n$.

The $T$ matrix in Eq.~(\ref{eq:teft}) is generated, order by order in the power
counting, by the $NN$ potential $V$ from
iterations of it in the Lippmann-Schwinger (LS) equation,
\begin{equation}
V+V\, G_0\, V+V\, G_0 \, V\, G_0 \, V+\dots \ ,
\label{eq:lse}
\end{equation}
where $G_0$ denotes the free two-nucleon propagator.
If we assume that,
\begin{equation}
\langle N^\prime N^\prime |V|NN\rangle=\sum_n V^{(n)}\ ,
\end{equation}
with $V^{(n)}$ of order $Q^n$, it is possible to assign to any $T^{(n)}$ the terms
in the LS equation that are of the same order. Generally in term like
$[V^{(m)}G_0V^{(n)}]$ is of order $Q^{m+n+1}$ because $G_0$ is of order
$Q^{-2}$ and the implicit loop integration brings a factor $Q^3$
(for a more detailed discussion, see Ref.~\cite{Pastore11}).

In our case the two nucleons interact via a PC potential plus a very small
TRV component. The $\chi$EFT Lagrangian implies
the following expansion in powers
of $Q$ for $T= T_{PC} + T_{TRV}$:
\begin{eqnarray}
 T_{PC}&=&T^{(0)}_{PC}+T^{(1)}_{PC}+T^{(2)}_{PC}+\ldots ,\label{tpc}\\
 T_{TRV}&=&T^{(-1)}_{TRV}+T^{(0)}_{TRV}+T^{(1)}_{TRV}+\ldots .\label{ttrv}
\end{eqnarray}
Assuming that the potential $V=V_{PC}+V_{TRV}$ have a similar expansion,
\begin{eqnarray}
 V_{PC} & = & V_{PC}^{(0)}+V_{PC}^{(1)}+V_{PC}^{(2)}+\dots \\
 V_{TRV} & = & V_{TRV}^{(-1)}+V_{TRV}^{(0)}+V_{TRV}^{(1)}+\dots ,
\end{eqnarray}
we can match order by order the $T$ and the terms in the LS equation obtaining
the form definition of the TRV $NN$ potential from the scattering amplitude,
\begin{eqnarray}
 V_{TRV}^{(-1)} & = & T_{TRV}^{(-1)}\ ,\label{eq:vtrvml}\\
 V_{TRV}^{(0)}  & = &  T_{TRV}^{(0)}-\left[V_{TRV}^{(-1)}G_0 V_{PC}^{(0)}\right]
                     -\left[V_{PC}^{(0)}G_0 V_{TRV}^{(-1)}\right]\ ,\label{eq:vtrv0}\\
 V_{TRV}^{(1)}  & = &  T_{PV}^{(1)}-\left[V_{TRV}^{(0)}G_0 V_{PC}^{(0)}\right]
               -\left[V_{PC}^{(0)}G_0 V_{TRV}^{(0)}\right]\nonumber\\
          & - & \left[V_{TRV}^{(-1)}G_0 V_{PC}^{(1)}\right]-
                \left[V_{PC}^{(1)}G_0 V_{TRV}^{(-1)}\right]\nonumber\\
          & - & \left[V_{TRV}^{(-1)}G_0 V_{PC}^{(0)}G_0 V_{PC}^{(0)}\right]
             - \left[V_{PC}^{(0)}G_0 V_{TRV}^{(-1)}G_0 V_{PC}^{(0)}\right]
             \nonumber\\
         & - & \left[V_{PC}^{(0)}G_0 V_{PC}^{(0)}G_0 V_{TRV}^{(-1)}\right]
             \ .\label{eq:vtrv1}
\end{eqnarray}
The generalization for the $NNN$ TRV potential is straightforward.

\subsection{The $NN$ TRV potential}
We define the following momenta,
\begin{eqnarray}
  &&\bmK_j=(\bmp_j'+\bmp_j)/2\ , \quad
  \bmk_j=\bmp_j'-\bmp_j\ ,\label{eq:notjb1}
\end{eqnarray}
where $\bmp_j$ and $\bmp'_j$ are the initial and final momenta
of nucleon $j$. From the overall momentum
conservation $\bmp_1+\bmp_2=\bmp_1'+\bmp_2'$, we can define
$\bmk=\bmk_1=-\bmk_2$. We also define $\bmK=(\bmK_1-\bmK_2)/2$,
$\bmP=\bmp_1+\bmp_2=\bmK_1+\bmK_2$,
in this way is possible to write the TRV $NN$ potential as,
\begin{equation}
  V_{TRV}(\bmk,\bmK_1,\bmK_2)= V_{TRV}^{(c.m.)}(\bmk,\bmK)+ 
   V^{(\bmP)}_{TRV}(\bmk,\bmK)\ ,\label{eq:widetildev2}
\end{equation}
where the term $ V^{(\bmP)}_{TRV}(\bmk,\bmK)$ represents boost
corrections to $  V^{(c.m.)}_{TRV}(\bmk,\bmK)$~\cite{girlanda10}, the potential
in the center-of-mass (c.m.) frame.  Below we ignore these boost corrections
and provide expressions for $ V^{(c.m.)}_{TRV}(\bmk,\bmK)$ only.

\begin{figure}[t]
  \begin{center}
    \includegraphics[scale=.5]{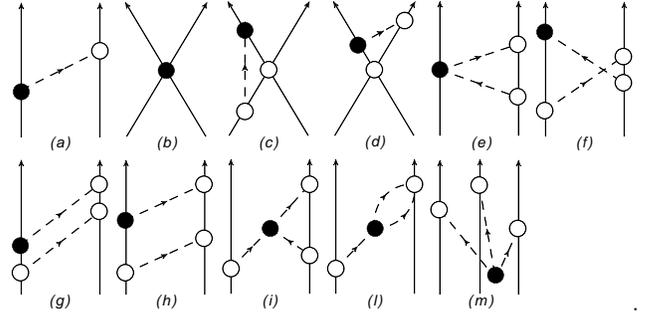}.
    \caption{   \label{fig:diagNN}
      Time-ordered diagrams contributing to the TRV
      potential (only a single time ordering is shown).
      Nucleons and pions are denoted by solid and dashed lines, respectively. The
      open (solid) circle represents a PC (TRV) vertex.}
  \end{center}
\end{figure}

The diagrams that give contribution to the $NN$ TRV potential are shown in
Fig.~\ref{fig:diagNN}.
We do not consider diagrams which give contribution only to the
renormalization of the LECs. In this section we write the final expression of
the $NN$ TRV potential $ V^{(c.m.)}_{TRV}$ as,
\begin{eqnarray}
    V^{(c.m.)}_{TRV}&=&
   V^{ ({\rm OPE})}_{TRV} +
   V^{ ({\rm TPE})}_{TRV} +
   V^{ ({\rm 3\pi,0})}_{TRV} +
   V^{ ({\rm 3\pi,1})}_{TRV} \nonumber\\
   &&+  V^{ ({\rm RC})}_{TRV} +
   V^{ ({\rm 3\pi,RC})}_{TRV} +
   V^{ ({\rm CT})}_{TRV} \ ,\label{eq:dec}
\end{eqnarray}
namely as a sum of terms due to one-pion exchange (OPE), two-pion exchange
(TPE), three-pion exchange at NLO (3$\pi$,0) and at N2LO (3$\pi$,1),
relativistic corrections derived from the OPE (RC) and from the $3\pi$-exchange
($3\pi$,RC),
and contact contributions (CT). From now on we define also
$g_0^*=g_0+g_2/3$ (see Appendix~\ref{app:pot} for details).
The OPE term is the contribution of diagram (a) of order $Q^{-1}$ (LO) and it reads,
\bgroup
\arraycolsep=0.7pt
\begin{eqnarray}
  \vtrv^{(\rm OPE)}
  &=&\frac{\overline{g}_A\overline{g}_0^*}{2\overline{f}_\pi}(\tone\cdot\ttwo)\frac{i\bmk\cdot
    (\sone-\stwo)}{\omk^2}\nonumber\\
  &+&
  \frac{\overline{g}_A\overline{g}_1}{4\overline{f}_\pi}\Big[(\tau_{1z}+\tau_{2z})
    \frac{i\bmk\cdot(\sone-\stwo)}{\omega_k^2}\nonumber\\
    &&\qquad
    +(\tau_{1z}-\tau_{2z})\frac{i \bmk\cdot(\sone+\stwo)}{\omega_k^2}\Big]
  \nonumber \\
  &+&\frac{\overline{g}_A\overline{g}_2}{6\overline{f}_{\pi}}
  (3\tau_{1z}\tau_{2z}-\tone\cdot\ttwo)\frac{i\bmk\cdot
    (\sone-\stwo)}{\omk^2}\ ,\nonumber\\
  \label{eq:isotope}
\end{eqnarray}
\egroup
where there are an isoscalar, an
isovector and an isotensor components and $\overline{g}_0^*=\overline{g}_0+\overline{g}_2/3$.
The coupling constants $\overline{g}_A/\overline{f}_{\pi}$,
$\overline{g}_0^*$, $\overline{g}_1$, $\overline{g}_2$ are renormalized
coupling constants, having reabsorbed the various infinites generated by loops and
given as combinations
of the ``bare'' LECs entering the Lagrangian. The expression for
$\overline{g}_A/\overline{f}_{\pi}$ is the same as reported in Ref.~\cite{viviani14}. The
LECs $g_{S1}^{(2)}$, $g_{S2}^{(2)}$, etc. enter in
Eq.~(\ref{eq:isotope}) through the renormalized constants $\overline{g}_0$, etc. (see
Appendix~\ref{app:pot} for more details). Note that, as mentioned before, the
correct expressions of the renormalized constants in term of the bare ones
should contain also the contributions of additional diagrams not
considered here (see Ref.~\cite{viviani14} for the procedure
to be followed for the PC and PV potentials, respectively).

The TPE term comes from the not singular contribution of panels (e)-(h) of Fig.
~\ref{fig:diagNN}, taking also into account the subtracting
terms given in Eq.~(\ref{eq:vtrv1}).
As discussed in Appendix~\ref{app:pot} this term has no isovector component,
in agreement with the result reported in~\cite{CM11}.
Therefore,
as reported in Eqs.~(\ref{eq:pote}) and~(\ref{eq:box}), we have,
\bgroup
\arraycolsep=1.0pt
\begin{align}
  \vtrv^{(\rm TPE)}&=\frac{\ga{g}_0^*}{\fp\Lx^2}
   \tone\cdot\ttwo \ i\bmk\cdot(\sone-\stwo)\ L(k)\nonumber\\
  &+\frac{\ga^3 {g}_0^*}{\fp\Lx^2}\ \tone\cdot\ttwo\  i\bmk\cdot(\sone-\stwo)
  \ (H(k)-3L(k))\nonumber\\
  &-\frac{\ga {g}_2}{3\fp\Lx^2}
   (3\tau_{1z}\tau_{2z}-\tone\cdot\ttwo)
   \ i\bmk\cdot(\sone-\stwo)\ L(k)\nonumber\\
   &-\frac{\ga^3 {g}_2}{3\fp\Lx^2}(3\tau_{1z}\tau_{2z}-\tone\cdot\ttwo)\nonumber\\
   &\qquad\qquad\times i\bmk\cdot(\sone-\stwo)\ 
   (H(k)-3L(k))\ ,
\end{align}
\egroup
where the loop functions $L(k)$ and $H(k)$ are defined in Eqs.~(\ref{eq:sL}) and
~(\ref{eq:H}). In this and the following NLO and N2LO terms 
the bare coupling constants $\ga$, $\fp$, $g_0^*$, $\ldots$ can be safely replaced by the corresponding
physical (renormalized) values.

The $3\pi$-exchange term gives a NLO contribution through the diagram (i) of
Fig.~\ref{fig:diagNN},
\bgroup
\arraycolsep=1.0pt
\begin{align}
  \vtrv^{(3\pi,0)}=&-\frac{5\ga^3\deltatre M}{4\fp\Lx^2}\pi
  \Big[(\tau_{1z}+\tau_{2z})
    \frac{i\bmk\cdot(\sone-\stwo)}{\omega_k^2}\nonumber\\
    &+
    (\tau_{1z}-\tau_{2z})\frac{i \bmk\cdot(\sone+\stwo)}{\omega_k^2}\Big]
  \nonumber\\
  &\qquad\times\Big(\big(1-\frac{2\mp^2}{s^2}\big)s^2A(k)+\mp\Big)\label{eq:3pnlo}\ ,
\end{align}
\egroup
where $A(k)$ is defined in Eq.~(\ref{eq:A}).
The expression obtained in Eq.~(\ref{eq:3pnlo}) is in agreement with the
expression derived in Refs.~\cite{JV13,JB150}.

The diagram in panel (l) of Fig.~\ref{fig:diagNN} contributes to
$\vtrv^{(3\pi)}$ at N2LO.
The expression for this diagram is derived in Appendix~\ref{app:pot},
here we report the final result only,
\bgroup
\arraycolsep=1.0pt
\begin{align}
  &\vtrv^{(3\pi,1)}=\frac{5\ga\deltatre Mc_1}{2\fp\Lx^2}
      \Big[(\tau_{1z}+\tau_{2z})i\bmk\cdot(\sone-\stwo)\nonumber\\
        &\qquad+(\tau_{1z}-\tau_{2z})i \bmk\cdot(\sone+\stwo)\Big]
       \ 4\frac{\mp^2}{\omega_k^2}L(k)\nonumber\\
       &\qquad-\frac{5\ga\deltatre Mc_2}{6\fp\Lx^2}
       \Big[(\tau_{1z}+\tau_{2z})i\bmk\cdot(\sone-\stwo)\nonumber\\
         &\qquad+(\tau_{1z}-\tau_{2z}) i \bmk\cdot(\sone+\stwo)\Big]
       \Big(2L(k)+6\frac{\mp^2}{\omega_k^2}L(k)\Big)\nonumber\\
       &\qquad-\frac{5\ga\deltatre Mc_3}{4\fp\Lx^2}
       \Big[(\tau_{1z}+\tau_{2z})i\bmk\cdot(\sone-\stwo)\nonumber\\
         &\qquad+(\tau_{1z}-\tau_{2z})i \bmk\cdot(\sone+\stwo)\Big]
       \Big(3L(k)+5\frac{\mp^2}{\omega_k^2}L(k)\Big)\label{eq:3pc3}\ .
\end{align}
\egroup
Note in Eq.~(\ref{eq:3pc3}) the presence of the
$c_1$, $c_2$ and $c_3$ LECs, which belong to the PC sector.
In Eqs.~(\ref{eq:3pnlo}) and~(\ref{eq:3pc3}) the $\deltatre$ is the
renormalized LEC to the relative order of the expressions.

The $\vtrv^{ ({\rm RC})}$ term of the potential takes into account contributions
from the RC of the vertices in the OPE of panel (a),
\begin{eqnarray}
  \vtrv^{(\rm RC)}&=&\frac{\ga {g}_0^*}{8\fp M^2}
  \tone\cdot\ttwo\ \frac{1}{\omk^2}\nonumber\\
  &&\times\Big[-\frac{i}{2} \left(8K^2+k^2\right)\bmk\cdot(\sone-\stwo)\nonumber\\
   &&+\bmk\cdot\sone(\bmk\times\bmK)\cdot\stwo
    -\bmk\cdot\stwo(\bmk\times\bmK)\cdot\sone\Big]\nonumber\\
  &+&\frac{\ga {g}_1}{16\fp M^2}\ \frac{1}{\omk^2}\nonumber\\
  &&\times\Big\{(\tau_{1z}-\tau_{2z})
  \Big[-\frac{i}{2} \left(8K^2+k^2\right)\bmk\cdot(\sone+\stwo)\nonumber\\
   &&+\bmk\cdot\sone(\bmk\times\bmK)\cdot\stwo
    +\bmk\cdot\stwo(\bmk\times\bmK)\cdot\sone\Big]\nonumber\\
  &&+(\tau_{1z}+\tau_{2z})
  \Big[-\frac{i}{2} \left(8K^2+k^2\right)\bmk\cdot(\sone-\stwo)\nonumber\\
   &&+\bmk\cdot\sone(\bmk\times\bmK)\cdot\stwo
    -\bmk\cdot\stwo(\bmk\times\bmK)\cdot\sone\Big]\Big\}\nonumber\\
  &+&\frac{\ga {g}_2}{24\fp M^2}(3\tau_{1z}\tau_{2z}-\tone\cdot\ttwo)
  \ \frac{1}{\omk^2}\nonumber\\
  &&\times\Big[-\frac{i}{2} \left(8K^2+k^2\right)\bmk\cdot(\sone-\stwo)\nonumber\\
   &&+\bmk\cdot\sone(\bmk\times\bmK)\cdot\stwo
    -\bmk\cdot\stwo(\bmk\times\bmK)\cdot\sone\Big]\label{eq:rco2}\ ,\nonumber\\
\end{eqnarray}
where, as for the OPE, we have an isoscalar, an isovector and an isotensor term.
Also the diagram of Fig.~\ref{fig:diagNN} (i) gives a contribution to the
at N2LO, both from the RC of the vertices, and from NLO in the pion propagators
(see Appendix~\ref{app:pot}). The final expression we obtain is
\begin{align}
  \vtrv&^{(\rm 3\pi,RC)}=-\frac{5\ga^3\deltatre}{16\fp\Lx^2}
  \Big[(\tau_{1z}+\tau_{2z})i\bmk\cdot(\sone-\stwo)\nonumber\\
  &+(\tau_{1z}-\tau_{2z})i \bmk\cdot(\sone+\stwo)\Big]\nonumber\\
  &\times\Big(\frac{25}{6}L(k)-\frac{7}{2}\frac{\mp^2}{\omega_k^2}L(k)
  +2\frac{\mp^2}{\omega_k^2}H(k)\Big)\nonumber\\
  &-\frac{25\ga^3\deltatre}{12\fp\Lx^2}\frac{1}
  {\omega_k^2}\ \Big\{(\tau_{1z}+\tau_{2z})
    \Big[\bmk\cdot\sone(\bmk\times\bmK)\cdot\stwo\nonumber\\
    &+\bmk\cdot\stwo(\bmk\times\bmK)\cdot\sone\Big]
  +(\tau_{1z}+\tau_{2z})\nonumber\\
  &\times\Big[\bmk\cdot\sone(\bmk\times\bmK)\cdot\stwo
    -\bmk\cdot\stwo(\bmk\times\bmK)\cdot\sone\Big]\Big\}\label{eq:rc3p}\ ,
  \nonumber\\
\end{align}
where in Eqs.~(\ref{eq:rco2}) and~(\ref{eq:rc3p}) the TRV coupling constants
are renormalized to order $Q^2$.

Last, the potential $V^{ ({\rm CT})}_{TRV}$, derived from the $NN$
contact diagrams (b) of Fig.~\ref{fig:diagNN}, reads
\begin{eqnarray}
  V^{\text{(CT)}}_{TRV}&=&\frac{1}{\Lambda_\chi^2f_\pi}\big\{\overline{C}_1\ i\bmk\cdot
      \left(\sone-\stwo\right)\nonumber\\
      &+&\overline{C}_2\ i\bmk\cdot
      \left(\sone-\stwo\right)\tone\cdot\ttwo\nonumber\\
      &+&{\overline{C}_3\over 2}\ 
      \big[i\bmk\cdot\left(\sone-\stwo\right)\left(\tau_{z1}+\tau_{z2}\right)
        \nonumber\\
        &&-i\bmk\cdot\left(\sone+\stwo\right)
        \left(\tau_{z1}-\tau_{z2}\right)\big]\nonumber\\
      &+&{\overline{C}_4\over 2}\ 
      \big[i\bmk\cdot\left(\sone-\stwo\right)\left(\tau_{z1}+\tau_{z2}\right)
        \nonumber\\
        &&+i\bmk\cdot\left(\sone+\stwo\right)
        \left(\tau_{z1}-\tau_{z2}\right)\big]\nonumber\\
      &+&\overline{C}_5\ i\bmk\cdot
      \left(\sone-\stwo\right)\left(3\tau_{z1}\tau_{z2}-\tone\cdot\ttwo\right)
      \big\}\label{eq:ct}\ .
\end{eqnarray}
Above $\overline{C}_1$, $\overline{C}_2$, $\overline{C}_3$, $\overline{C}_4$ and $\overline{C}_5$ are
renormalized LECs since they have reabsorbed various singular terms coming
from the TPE diagrams, 3$\pi$ diagrams and the relativistic corrections.
Note that it is possible to write ten operators which can enter
$V_{TRV}^{(\text{CT})}$ at order $Q$ but only five of them are independent.
In this work we have chosen to write the operators in term of $\bmk$,
so that the $r$-space version of $V_{TRV}^{(\text{CT})}$
will assume a simple local form with no gradients.
We want also to remark that the $C_2$, $C_4$
and $C_5$ LECs are needed in order to reabsorb the divergences
coming from the TPE and $3\pi$ exchange diagrams.

In the calculation of the EDM in Sec.~\ref{sec:res},
the configuration space version of the potential is needed.
This formally follows from 
\bgroup
\arraycolsep=1.0pt
\begin{eqnarray} 
 \!\!\!\!\!\!\!\!\!\!\!\! \langle \bmr_1'\bmr_2'|V|\bmr_1\bmr_2\rangle&=&
  \delta (\bmR-\bmR') \int {d^3k\over (2\pi)^3} {d^3K\over (2\pi)^3}
  \nonumber\\ 
 &\times& e^{i(\bmK+\bmk/2)\cdot\bmr'} V(\bmk,\bmK)
    e^{-i(\bmK-\bmk/2)\cdot\bmr}\ ,
    \label{eq:vrsp}
\end{eqnarray}
\egroup
where $\bmr=\bmr_1-\bmr_2$ and $\bmR=(\bmr_1+\bmr_2)/2$,
and similarly for the primed variables.
In order to carry out the Fourier transforms above, 
the integrand is regularized by including a cutoff of the form
\begin{equation}
   C_\Lambda(k)={\rm e}^{-(k/\Lambda)^4}\ ,\label{eq:cutoff}
\end{equation}
where the cutoff parameter $\Lambda$ is taken in the range
450--550 MeV.  With such a choice the  $V^{ ({\rm OPE})}_{TRV}$,
   $V^{ ({\rm TPE})}_{TRV}$,
   $V^{ ({\rm 3\pi,0})}_{TRV}$,
   $V^{ ({\rm 3\pi,1})}_{TRV}$, 
 and $V^{ ({\rm CT})}_{TRV}$
components of the resulting potential are local, i.e.,
$\langle \bmr_1'\bmr_2'|V|\bmr_1\bmr_2\rangle=
\delta (\bmR-\bmR')\, \delta (\bmr-\bmr') V(\bmr)$, while
the RC component contains mild non-localities
associated with linear and quadratic terms in the
relative momentum operator $-i \bmna$.
Explicit expressions for all these
components are listed in Appendix~\ref{app:rpotNN}.

\subsection{The $NNN$ TRV potential}\label{sec:NNNtrv}

The $3\pi$ TRV vertex gives rise to a three body contribution through the
diagram (m) in Fig.~\ref{fig:diagNN}.
The lowest contribution appears at NLO while at
N2LO the various time ordering cancel out (see Appendix~\ref{app:pot}).  
The final expression for the NLO of the $NNN$ TRV potential is,
\begin{eqnarray}
  \vtrv^{\rm{NNN}}&=&\frac{\deltatre g_A^3 M}{4\fp^3}(\tone\cdot\ttwo\ \tau_{3z}+
  \tone\cdot\ttre\ \tau_{2z}+\ttwo\cdot\ttre\ \tau_{1z})\nonumber\\
  &&\times\frac{(i\bmk_1\cdot\sone)\ (i\bmk_2\cdot\stwo)\ (i\bmk_3\cdot\stre)}
                       {\omega^2_{k_1}\omega^2_{k_2}\omega^2_{k_3}}\ ,
                       \label{eq:NNNpot}
\end{eqnarray}
which is in agreement with the expression of Ref.~\cite{JV13,JB150}.

Also in this case we need the Fourier transform of the potential.
Using the overall momentum conservation
$\bmp_1+\bmp_2+\bmp_3=\bmp_1'+\bmp_2'+\bmp_3'$, which give us that
$\bmk_3=-\bmk_1-\bmk_2$, and defining $\bmQ=\bmk_1+\bmk_2$ and
$\bmq=\bmk_1-\bmk_2$ the Fourier transform becomes,
\begin{align} 
 \langle \bmr_1'\bmr_2'\bmr_3'|& V|\bmr_1\bmr_2\bmr_3\rangle=
 \delta (\bmr_1-\bmr_1')\delta (\bmr_2-\bmr_2')\delta (\bmr_3-\bmr_3')
 \nonumber\\ 
 &\times\int {d^3q\over (2\pi)^3} {d^3Q\over (2\pi)^3}
 V(\bmq,\bmQ)
    e^{-i(\bmq/2)\cdot\bmx_2}\,e^{-i\bmQ\cdot\bmx_1}\ ,
    \label{eq:vrnnn}\nonumber\\
\end{align}
where $\bmx_1=\bmr_2-\bmr_1$ and $\bmx_2=\bmr_3-(\bmr_1+\bmr_2)/2$.
Note that with this choice of $\bmQ$ and $\bmq$, the form of the Jacobi
vectors appear automatically.
We use the regularization function reported in Eq.~(\ref{eq:cutoff}), where
we replace $k$ with $Q$, namely,
\begin{equation}
   C_\Lambda(Q)={\rm e}^{-(Q/\Lambda)^4}\ ,\label{eq:cutoffnnn}
\end{equation}
which gives a local form of the $NNN$ TRV potential. A complete description of
how to carry out the integration in the Fourier transform is given in
Appendix~\ref{sec:NNNr}.

\section{Results}\label{sec:res}

In this section, we report results for the EDM of $\det$, $\tri$ and $\hel$.
Hereafter, we do not use the barred notation for the
renormalized LECs but all the coupling constant
must be considered as renormalized.
The calculations are based on the TRV $NN$ potential derived
in the previous section (and summarized in Appendix~\ref{app:pot}) and on the
(strong interaction) PC $NN$ potential obtained by Entem and Machleidt
at next-to-next-to-next-to-next-to-leading order (N4LO)~\cite{DR15}. These
potentials are regularized with a cutoff function depending on a
parameter $\Lambda$; its functional form, however, is different from the
adopted here for $V_{TRV}$. Below we consider the versions with
$\Lambda=450$ MeV, $500$ MeV, and $550$ MeV.
The calculations of $\tri$ and $\hel$ EDMs also include the $NNN$ TRV nuclear potential
derived in Sec.~\ref{sec:NNNtrv} and the PC $NNN$ potential derived in $\chi$EFT
at N2LO. As for the $NN$ PC potential, it depends on a cutoff parameter
$\Lambda$ which is chosen to be consistent with those in the PC and TRV
$NN$ potentials. The three-nucleon PC potential depends, in addition,
on two unknown LECs, denoted as $c_D$ and $c_E$ and also
on the LECs $c_1$, $c_3$ and $c_4$. In this work, we use
the values reported in Table III of Ref.~\cite{LE18}. The $\vtrv^{(3\pi,1)}$ term
of the TRV potential
depends on the LECs $c_1$, $c_2$ and $c_3$ which are taken from
Table II of Ref.~\cite{DR15} and summarized here 
in Table~\ref{tab:cecd}.
\begin{table}[h]
\begin{center}
\begin{tabular}{lccc}
  \hline
  \hline
PC interactions & $c_1$ & $c_2$ & $c_3$ \\
\hline
N2LO  & -0.74 &   -  & -3.61 \\
N3LO  & -1.07 & 3.20 & -5.32 \\
N4LO  & -1.10 & 3.57 & -5.54 \\
\hline
\hline
\end{tabular}
\caption{ \label{tab:cecd}
  Values of the coefficients
  and $c_1$, $c_2$, $c_3$  in unit of
  GeV$^{-1}$ taken from Ref.~\cite{DR15}.}
\end{center}
\end{table}
In the following the values  $g_A=1.267$ and $\fp=92.4$ MeV are
adopted.

This section is organized as follows. In Sec.~\ref{sec:dedm}, we present
the general expression for the EDM operator and the results for
the deuteron EDM, while in Sec.~\ref{sec:3Hedm} we present the calculation of the
$\tri$ and $\hel$ EDMs.

\subsection{Deuteron EDM}\label{sec:dedm}
The EDM operator $\EDM$ is composed by two parts,
\begin{equation}
  \EDM=\EDM_{\rm PC}+\EDM_{\rm TRV}.
\end{equation}
The $\EDM_{\rm PC}$ receives contribution at LO from the nuclear EDM polarization operator
\begin{equation}
  \EDM_{\rm PC}=e\sum_i\frac{1+\tau_z(i)}{2}\boldsymbol{r}_i,
  \label{eq:dpc}
\end{equation}
where $e>0$ is the electric unit charge, $\tau_z(i)$ and $\bmr_i$ are the
$z$ component of the isospin and the position of the i-th particle.
The $\EDM_{\rm TRV}$ LO contribution comes from the intrinsic nucleon EDM,
\begin{equation}
  \EDM_{\rm TRV}=\frac{1}{2}\sum_i\left[(d_p+d_n)+(d_p-d_n)\tau_z(i)
    \right]\boldsymbol{\sigma}(i)\ ,
  \label{eq:dtrv}
\end{equation}
where $d_p$ and $d_n$ are the EDM of proton and neutron respectively and
$\boldsymbol{\sigma}(i)$ is the spin operator which act on the i-th particle.
As discussed in Refs.~\cite{JV11b,JB13}
both the $\EDM_{\rm PC}$ and $\EDM_{\rm TRV}$
receive contributions from transition currents at N2LO.
Of course, a complete treatment
of the EDM up to N2LO needs to take care of them but hereafter
we neglect their contribution, showing only the effects of N2LO
TRV potential.
In future work we plan to include the N2LO current contributions in the
calculations.

The nucleon EDM of an $A$ nucleus can be expressed as
\begin{eqnarray}
  d^A & = & \bra \psi^A_{+} | \hat{D}_{\rm TRV} | \psi^A_{+} \ket
  +2\, \bra \psi^A_{+} | \hat{D}_{\rm PC} | \psi^A_{-} \ket
  \nonumber\\
  & \equiv & d_{\rm TRV}^A+d_{\rm PC}^A\ ,
\end{eqnarray}
where $|\psi^A_{+}\ket$ $(|\psi^A_{-}\ket)$
is defined to be the even-parity (odd-parity) component of the wave function.
In general, due to the smallness of the LECs, the EDM  can be
expressed as linear on the TRV LECs,
\begin{eqnarray}
  d^A_{\rm TRV}&=&d_pa_p+d_na_n\\
  d^A_{\rm PC}&=&g_0a_0+g_1a_1+g_2a_2+\Delta_3 a_\Delta\nonumber\\
  &+&C_1A_1+C_2A_2+C_3A_3+C_4A_4+C_5A_5\ ,
  \label{eq:da}
\end{eqnarray}
where the $a_i$ for $i=0,1,2$, $a_\Delta$, $A_i$ for $i=1,\dots,5$  and
$a_p$, $a_n$ are
numerical coefficients independent on the LECs values
(however, they do depend on the cutoff $\Lambda$
in the PC and TRV chiral potentials).

We evaluate also the theoretical errors associated with the chiral expansion
of the nuclear potential. We express the error on the numerical coefficients
for the deuteron as,
\begin{equation}
  \left(\delta a_i\right)^2=\left(\delta a_i^{\text{PC}}\right)^2
  +\left(\delta a_i^{\text{TRV}}\right)^2\ ,
  \label{eq:errd}
\end{equation}
where $\delta a_i^{\text{PC}}$ is the error associated to the chiral expansion
of the PC potential and $\delta a_i^{\text{TRV}}$ the error associated
with the chiral expansion of the TRV potential.
Both the contributions were evaluated following the
prescriptions of Epelbaum {\it et al.}~\cite{EE15}
where as reference momentum in the calculation of the errors
we used the mass of the pion. 
It is straightforward
to understand that the errors are dominated by the TRV part because we are using
the N2LO potential for it and the N4LO potential for the PC part.

The coefficients for the deuteron, evaluated with the N4LO PC potential
and the N2LO TRV potential and the associated errors
for the three different choices of the cutoff
parameters, are given in Table~\ref{tab:2hdpc}.
The coefficients $a_p$ and $a_n$ multiplying the intrinsic neutron and
proton EDM, as already pointed out first in Ref.~\cite{NY15} and then
in Ref.~\cite{JB151}, are given by,
\begin{equation}
  a_n=a_p=\left(1-\frac{3}{2}P_D\right)\, ,
\end{equation}
where $P_D$ is the percentage of D-wave present in the deuteron wave function.
The values of $a_n$ and $a_p$ obtained
using the Entem and Machleidt at N4LO for three different choices of the cut off
are reported in Table~\ref{tab:2hdpc}.
The operator $\EDM_{\rm PC}$ for the deuteron reduces to
$\EDM_{\rm PC}=(\tau_z(1)-\tau_z(2))\bmr$ therefore $d^2_{\rm PC}$ in Eq.~(\ref{eq:da})
receives contribution only from the component $^3P_1$ which has $T=1$. This component
is generated by the TRV potential components proportional to $(\sone+\stwo)$,
namely those proportional to $g_1$, $\Delta_3$, $C_3$ and $C_4$.
\begin{table}[t]
\begin{center}
\begin{tabular}{lccc}
    \hline
    \hline
    $\Lambda$(MeV) & $450$ & $500$ & $550$\\
    \hline
    $a_n(a_p)$ & $\m0.934\pm0.001$ & $\m0.939\pm0.001$ & $\m0.938\pm0.001$\\
    $a_1$      & $\m0.192\pm0.006$ & $\m0.197\pm0.004$ & $\m0.194\pm0.003$\\ 
    $a_\Delta$ & $ -0.306\pm0.174$ & $ -0.341\pm0.153$ & $ -0.349\pm0.137$\\ 
    $A_3$      & $\m0.013\pm0.004$ & $\m0.013\pm0.004$ & $\m0.013\pm0.004$\\ 
    $A_4$      & $ -0.013\pm0.004$ & $ -0.013\pm0.004$ & $ -0.013\pm0.004$\\
    \hline
    \hline
\end{tabular}
\caption{ \label{tab:2hdpc}
  Values of the deuteron coefficients $a_n$ and $a_p$ in units of
  $d_p$ and $d_n$ and of  $a_1$, $a_\Delta$, $A_3$, $A_4$
  in units of $e$ fm
  for the three different choices of cutoff parameters $\Lambda$.}
\end{center}
\end{table}

The coefficient $a_\Delta$ can be written as,
\begin{equation}
  a_\Delta=a_\Delta(0)+\big(c_1a_\Delta(1)+c_2a_\Delta(2)+c_3a_\Delta(3)\big)+
  a_\Delta(RC)\ ,\label{eq:adelta}
\end{equation}
where the first term comes from $V_{TRV}^{(3\pi,0)}$,
the term in parenthesis from
$V_{TRV}^{(3\pi,1)}$,  where the LECs $c_1$, $c_2$ and $c_3$ appear,
and the last term from $V_{TRV}^{(3\pi,{\rm RC})}$. In Table~\ref{tab:2had}
we report the values of the coefficients $a_\Delta(i)$ evaluated using the
N4LO PC potential for the various cut-off.
\begin{table}[t]
\begin{center}
\begin{tabular}{lccc}
    \hline
    \hline
    $\Lambda$(MeV) & $450$ & $500$ & $550$\\
    \hline
    $a_\Delta(0) $ & $ -0.872$ & $ -0.894$ & $ -0.894$\\
    $a_\Delta(1) $ & $\m0.117$ & $\m0.120$ & $\m0.120$\\ 
    $a_\Delta(2) $ & $ -0.119$ & $ -0.119$ & $ -0.117$\\ 
    $a_\Delta(3) $ & $ -0.209$ & $ -0.207$ & $ -0.203$\\ 
    $a_\Delta(RC)$ & $ -0.042$ & $ -0.037$ & $ -0.032$\\
    \hline
    \hline
\end{tabular}
\caption{ \label{tab:2had}
  Values of the various components of the deuteron
  coefficients $a_\Delta$ as given in Eq.~(\ref{eq:adelta})
  in units of $e$ fm evaluated using the N4LO PC potential
  for the three different choices of cutoff parameters $\Lambda$.}
\end{center}
\end{table}
We observe that, taking individually the coefficients which multiplies the
LECs $c_i$, their correction to the NLO is between $13-23\%$ which is in line
with what we expect from the chiral expansion.
However, using the values of $c_i$ given in Table~\ref{tab:cecd}
the total correction of $V^{(3\pi,1)}_{TRV}$ is about $66\% $,
which is very large compared to
what we expect adding a new order in the potential.
Therefore, the value of $a_\Delta$ is very sensitive
to the $V_{TRV}^{(3\pi,1)}$ component of the potential and on the choice
of the values of the constant $c_i$ and in particular of $c_3$.
This is reflected also in the large error associated with this coefficient
reported in Table~\ref{tab:2hdpc}.
On the other hand, the $V_{TRV}^{(3\pi,{\rm RC})}$ contribution is about $\sim4\%$
as expected for a relativistic corrections.
We notice also that the values of the coefficients for the three different
cut-off are compatible within the error bars.

We now compare our results with the values reported in Ref.~\cite{JB151}
where the Authors used the same TRV potential at NLO with
a N2LO PC potential with three-body forces~\cite{EE09,EE05}.
In Table~\ref{tab:2hdcomp} we compare our results obtained with
our TRV potential up to NLO and N2LO with a cutoff $\Lambda=500$ MeV with the
ones reported in Ref.~\cite{JB151}.

\begin{table}[t]
\begin{center}
\begin{tabular}{lccccc}
  \hline
  \hline
TRV pot. &$a_n(=a_p)$& $a_1$ & $a_\Delta$ & $A_3$ & $A_4$\\
\hline
Ref.~\cite{JB151}(NLO)& $0.939$& $0.183$ & $-0.748$ & $\m0.006$& $-0.006$\\
This work (NLO)  & $0.939$& $0.200$ & $-0.893$ & $-$      & $-$\\
This work (N2LO) & $0.939$& $0.197$ & $-0.341$ & $0.013$& $-0.013$\\
\hline
\hline
\end{tabular}
\caption{ \label{tab:2hdcomp}
  Comparison of the coefficients $a_1$, $a_\Delta$, $A_3$ and $A_4$
  for the deuteron with the result of Refs.~\cite{JB151}.
  To be notice that in this work we use $e>0$. For this work
  we report the calculation up to NLO and N2LO.}
\end{center}
\end{table}

In order to compare the values of the $A_i$ coefficient we divide the reported
values for $\frac{2(\hbar c)^3}{\Lx^2\fp}$
which permits to connect the two potentials.
As can be seen from Table~\ref{tab:2hdcomp}, our values at NLO
seem to be systematically larger compared to 
the values reported in Ref.~\cite{JB151}.
Even if we use a different PC potential, the reason should be
found in the different regularization function.
However, from a qualitative point of view there is a substantial agreement with
Ref.~\cite{JB151}. Similar agreement has been founded for $a_1$
with the result reported
in Refs.~\cite{CP04,NY15,YH13} while are smaller compare the results reported in
Refs.~\cite{JV11,IS08}.

\subsection{$\tri$ and $\hel$ EDMs}\label{sec:3Hedm}

In this section we report the results obtained for the EDM of the $\tri$ and
$\hel$. The wave functions of $\tri$ and $\hel$ have been obtained with
the hyperspherical harmonics (HH)~\cite{AK08,LE09} from the Hamiltonians N4LO/N2LO-500,
N4LO/N2LO-450 and N4LO/N2LO-550 discussed in Sec.~\ref{sec:res}. Moreover,
we evaluated also the
errors on the numerical coefficients due to the chiral expansion as,
\begin{equation}
  \left(\delta a_i\right)^2=\left(\delta a_i^{\text{PC}}\right)^2
  +\left(\delta a_i^{\text{TRV}}\right)^2+\left(\delta a_i^{\psi}\right)^2\ ,
  \label{eq:errh}
\end{equation}
where $\delta a_i^{\text{PC}}$ and $\delta a_i^{\text{TRV}}$ are the same as
in Eq.~(\ref{eq:errd}), while $\delta a_i^{\psi}$ is the error associated to the
numerical accuracy of the
3-body wave function which we estimated to be of the order of $\sim 1\%$.
The calculated values of the numerical coefficients
for the three choices of the cutoff with their associated  errors are reported in
Table~\ref{tab:3htot}.

The $a_\Delta$ can be written as,
\begin{eqnarray}
  a_\Delta&=&a_\Delta(0)+\big(c_1a_\Delta(1)+c_2a_\Delta(2)+c_3a_\Delta(3)\big)+
  \nonumber\\
  &&a_\Delta(RC)+a_\Delta(3N) ,\label{eq:adelta3}
\end{eqnarray}
where $a_\Delta(0)$, $a_\Delta(1)$, $a_\Delta(2)$, $a_\Delta(3)$ and $a_\Delta(RC)$
are defined as in Eq.~(\ref{eq:adelta}) while $a_\Delta(3N)$ represents the TRV 3-body
potential contribution. In Table~\ref{tab:3had}
we report the values of the coefficients $a_\Delta(i)$ evaluated using the
N4LO/N2LO-$\Lambda$ PC potential for the various cut-off.
\begin{table}[t]
\begin{center}
\begin{tabular}{lcccccc}
    \hline
    \hline
    \multicolumn{1}{c}{} & \multicolumn{3}{c}{$\tri$}& \multicolumn{3}{c}{$\hel$}\\
    $\Lambda$(MeV) & $450$ & $500$ & $550$ & $450$ & $500$ & $550$\\
    \hline
    $a_\Delta(0) $&$ -0.716$&$ -0.751$&$ -0.758$&$ -0.716$& $ -0.749$& $ -0.755$\\
    $a_\Delta(1) $&$\m0.093$&$\m0.098$&$\m0.099$&$\m0.093$& $\m0.098$& $\m0.099$\\
    $a_\Delta(2) $&$ -0.107$&$ -0.110$&$ -0.110$&$ -0.106$& $ -0.109$& $ -0.109$\\
    $a_\Delta(3) $&$ -0.194$&$ -0.198$&$ -0.198$&$ -0.192$& $ -0.196$& $ -0.196$\\
    $a_\Delta(RC)$&$ -0.048$&$ -0.046$&$ -0.042$&$ -0.048$& $ -0.044$& $ -0.041$\\
    $a_\Delta(3N)$&$ -0.202$&$ -0.190$&$ -0.205$&$ -0.193$& $ -0.180$& $ -0.196$\\
    \hline
    \hline
\end{tabular}
\caption{ \label{tab:3had}
  Values of the various components of $a_\Delta$ as given in Eq.~(\ref{eq:adelta3})
  for $\tri$ and $\hel$ in units of $e$ fm evaluated using the N4LO/N2LO PC potential
  for the three different choices of cutoff parameters $\Lambda$.}
\end{center}
\end{table}
The $a_\Delta(3N)$ give a correction to $a_\Delta(0)$  of the order
of the $\sim25\%$, which is in line with the chiral perturbation theory
prediction because both
the contributions appear at the same order. For completeness we report
in Table~\ref{tab:3Nconv} of Appendix~\ref{app:NNNconv}
the convergence pattern of this
contribution as function of the HH basis expansion.
For both $\tri$ and $\hel$,
using the values of the $c_i$ reported in Table~\ref{tab:cecd},
we have found a correction to $a_\Delta(0)$ due to  $\vtrv^{(3\pi,1)}$ of
 $\sim79\%$, and of order $\sim6\%$ due to $\vtrv^{(3\pi,{\rm RC})}$. While the
RC are in line with what we expect, the $\vtrv^{(3\pi,1)}$ corrections
have much more impact on the values of $a_\Delta$
than predicted by the chiral perturbation theory.
The large correction due to $\vtrv^{(3\pi,1)}$ that appears at N2LO
is also reflected in the large uncertainties associated with this coefficient.
Again, the estimated uncertainties depends critically on the adopted values
of $c_1$, $c_2$, and $c_3$ (see Table~\ref{tab:cecd}).

The contribution of the TPE diagrams to $a_0$ and $a_2$
are of the order
of $\sim45\%$ and $\sim40\%$ respectively which is larger than expected from the
chiral convergence and they are due mainly to
the box diagram in Fig.~\ref{fig:diagNN}.
On the other hand, the RC to
$a_0$, $a_1$ and $a_2$ are of the order of $\sim1-3\%$
perfectly consistent with the prediction of the chiral perturbation theory.
The effects of the PC $NNN$ potential on the values of all the coefficients
is around $\sim2\%$.
From Table~\ref{tab:3htot} it is possible also to observe that
the values of the numerical coefficients are mostly equal in modulus between
$\tri$ and $\hel$ except $a_p$ and $a_n$. Moreover, it is possible to observe
that the isovector terms have the same sign for $\tri$ and $\hel$.
The isovector part of the TRV potential depends on the third component
of the isospin, therefore 
the $|\psi^A_{-}\ket$ wave function component generated  
has a sign $-$ for  $\tri$ and a $+$ for $\hel$. However, in the $\EDM_{\rm PC}$ there is the $\tau_z(i)$ operator
which bring again a sign $-$ for $\tri$ and $+$ for $\hel$
giving in total a $+$.

In Table~\ref{tab:3hcomp} we compare our calculations at NLO for a value of the
cutoff $\Lambda=500$ MeV with the values reported
in Ref.~\cite{JB151}.
As can be seen inspecting the values of the coefficients evaluated at NLO,
there is a nice agreement with the results of Ref.~\cite{JB151}.
The numerical differences (which however are within the error bars reported
in Ref.~\cite{JB151}) must be searched in the different PC potential and
in the different regularization function in the TRV potential employed here.
On the other hand for $a_\Delta^{(3)}$, the pure three body part of
$a_\Delta$, the difference is of one order of magnitude
which can not be explained by the different regularization function used.
We found a good agreement for the values of $a_0$ and $a_1$
with the results obtained using a phenomenological potential
and the meson exchange TRV potential in Refs.~\cite{NY15,YH13}
while we obtain
smaller values compare to Refs.~\cite{JV11,IS08}.

\begin{figure}[t]
  \begin{center}
  \includegraphics[width=8cm,clip]{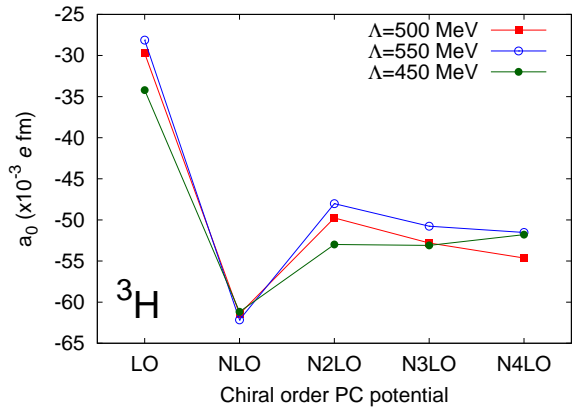}
  \caption{\label{fig:a0coeff} Values of the coefficient $a_0$ for
    the $\tri$ nucleus and 
    for the three choice of the cutoff when varying the chiral
    order of the PC potential.}
  \end{center}
\end{figure}

\begin{figure}[t]
  \begin{center}
  \includegraphics[width=8cm,clip]{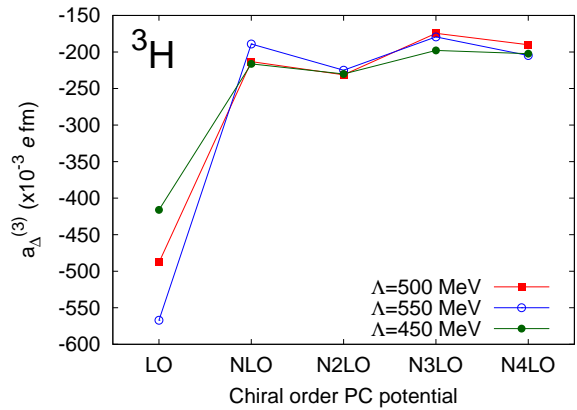}
  \caption{\label{fig:Dcoeff} The same as Fig.~\ref{fig:a0coeff} but for the
  coefficient $a_\Delta^{(3)}$.}
  \end{center}
\end{figure}

In the case of $\tri$ and $\hel$ we studied also the dependence of
the values of the coefficients on the chiral order of the PC potential.
As example, in Fig.~\ref{fig:a0coeff} and~\ref{fig:Dcoeff}
we show the values of the coefficient $a_0$
and $a_\Delta^{(3)}$ for $\tri$ evaluated using the N2LO TRV potential with different
chiral order of the PC potential and for the three different
choices of the cutoff.
In all the case we studied, the values evaluated with the N2LO, N3LO and N4LO
PC potential
differ by less than $5\%$ which confirm the robustness of the calculation.
\begin{widetext}
  \begin{center}
\begin{table}[t]
\begin{center}
\begin{tabular}{lcccccc}
  \hline
  \hline
\multicolumn{1}{c}{} & \multicolumn{3}{c}{$\tri$}& \multicolumn{3}{c}{$\hel$}\\
$\Lambda$ (MeV) & $450$ & $500$ & $550$ & $450$ & $500$ & $550$ \\
\hline
$a_n$      &$ -0.032\pm0.001$&$ -0.033\pm0.001$&$ -0.035\pm0.001$&$\m0.900\pm0.009$&$\m0.908\pm0.009$&$\m0.906\pm0.009$\\
$a_p$      &$\m0.901\pm0.009$&$\m0.909\pm0.009$&$\m0.907\pm0.009$&$ -0.032\pm0.001$&$ -0.033\pm0.001$&$ -0.034\pm0.001$\\
$a_0$      &$ -0.052\pm0.012$&$ -0.055\pm0.013$&$ -0.052\pm0.012$&$\m0.053\pm0.012$&$\m0.056\pm0.013$&$\m0.052\pm0.012$\\
$a_1$      &$\m0.147\pm0.005$&$\m0.154\pm0.004$&$\m0.155\pm0.003$&$\m0.148\pm0.005$&$\m0.155\pm0.004$&$\m0.155\pm0.003$\\
$a_2$      &$ -0.114\pm0.010$&$ -0.120\pm0.009$&$ -0.121\pm0.008$&$\m0.112\pm0.009$&$\m0.118\pm0.009$&$\m0.119\pm0.008$\\
$a_\Delta$ &$ -0.378\pm0.105$&$ -0.388\pm0.101$&$ -0.407\pm0.088$&$ -0.373\pm0.106$&$ -0.383\pm0.102$&$ -0.402\pm0.089$\\
$A_1$      &$\m0.005\pm0.001$&$\m0.006\pm0.001$&$\m0.006\pm0.001$&$ -0.005\pm0.002$&$ -0.006\pm0.002$&$ -0.006\pm0.001$\\
$A_2$      &$ -0.009\pm0.003$&$ -0.010\pm0.003$&$ -0.010\pm0.003$&$\m0.009\pm0.003$&$\m0.010\pm0.003$&$\m0.010\pm0.002$\\
$A_3$      &$ -0.008\pm0.002$&$ -0.008\pm0.002$&$ -0.008\pm0.002$&$ -0.008\pm0.003$&$ -0.008\pm0.002$&$ -0.008\pm0.002$\\
$A_4$      &$\m0.012\pm0.004$&$\m0.013\pm0.004$&$\m0.013\pm0.004$&$\m0.012\pm0.004$&$\m0.013\pm0.004$&$\m0.013\pm0.003$\\
$A_5$      &$ -0.021\pm0.006$&$ -0.022\pm0.006$&$ -0.022\pm0.006$&$\m0.020\pm0.006$&$\m0.022\pm0.006$&$\m0.022\pm0.005$\\
\hline
\hline
\end{tabular}
\caption{ \label{tab:3htot}
  Values of the numerical coefficients for $\tri$ and $\hel$ in units of $e$ fm
  ($a_n$ $(a_p)$ in units of $d_n$ ($d_p$)) for the three different choices of
  the cutoff.}
\end{center}
\end{table}
\begin{table}[t]
\begin{center}
\begin{tabular}{lcccccc}
  \hline
  \hline
\multicolumn{1}{c}{} & \multicolumn{3}{c}{$\tri$}& \multicolumn{3}{c}{$\hel$}\\
& This work  & This work  & Ref.~\cite{JB151}
& This work  & This work  & Ref.~\cite{JB151}\\
&  (NLO) &  (N2LO) & (NLO)
&  (NLO) &  (N2LO) & (NLO)\\

\hline
$a_n$      &$ -0.033$&$ -0.033$&$ -0.030$&$\m0.908$&$\m0.908$&$\m0.904$\\
$a_p$      &$\m0.909$&$\m0.909$&$\m0.918$&$ -0.033$&$ -0.033$&$ -0.029$\\
$a_0$      &$ -0.101$&$ -0.055$&$ -0.108$&$\m0.101$&$\m0.056$&$\m0.111$\\
$a_1$      &$\m0.158$&$\m0.154$&$\m0.139$&$\m0.158$&$\m0.155$&$\m0.139$\\
$a_2$      &$\m0.087$&$\m0.120$&  n.a.   &$\m0.086$&$\m0.118$& n.a.\\
$a_\Delta^{(2)}$ &$ -0.751$&$ -0.198$&$-0.598$&$ -0.749$&$ -0.202$&$-0.608$\\
$a_\Delta^{(3)}$ &$ -0.190$&$ -     $&$-0.017$&$ -0.180$&$ -     $&$-0.017$\\
$A_1$      &$-$&$\m0.006$&$\m0.005$&$-$&$ -0.006$&$ -0.005$\\
$A_2$      &$-$&$ -0.010$&$ -0.011$&$-$&$\m0.010$&$\m0.011$\\
$A_3$      &$-$&$ -0.008$&$ -0.005$&$-$&$ -0.008$&$ -0.005$\\
$A_4$      &$-$&$\m0.013$&$\m0.009$&$-$&$\m0.013$&$\m0.009$\\
$A_5$      &$-$&$ -0.022$&   n.a.  &$-$&$\m0.022$& n.a.     \\
\hline
\hline
\end{tabular}
\caption{ \label{tab:3hcomp}
  Comparison of the values of the coefficients for $\tri$ and $\hel$
  obtained for $\Lambda=500$ MeV with the results of Ref.~\cite{JB151}.
  $a_\Delta^{(2)}$ and $a_\Delta^{(3)}$ correspond respectively to the 2-body
  and 3-body contribution to $a_\Delta$.
  To be noticed that in this work we use $e>0$.}
\end{center}
\end{table}
\end{center}
\end{widetext}

\section{Conclusions}\label{sec:conc}

In this work we derived the TRV $NN$ and $NNN$ potential at N2LO.
In order to derive the potential we have considered the most generic Lagrangian
without any specific hypothesis for the TRV source.
With the derived potential, we have
calculated the EDM of $\det$, $\tri$ and $\hel$ investigating the
effect of the N2LO components. We have found that the sensitivity
of the light nuclei EDM  to the LEC $\Delta_3$ found at NLO, is well reduced by the
N2LO contribution which is a quite unexpected behavior inside the
chiral perturbation framework. We also checked the robustness of our
calculation, evaluating the EDM of the nuclei using different chiral orders
in the PC potential. The discrepancy between the use of the N2LO and the N4LO
PC potential is approximately $5\%$.

We have compared our results with the existing other values reported in literature
and in particular with the calculation of Ref.~\cite{JB151}.
We have found a substantial agreement with the results reported in
Ref.~\cite{JB151}
where the small numerical differences can be originated by the different
function used to regularize the potential.
We have found a qualitative agreement with those of Refs.~\cite{CP04,NY15,YH13}
while we have obtained smaller values compared to Refs.~\cite{JV11,IS08}.

Our results depend on eleven coupling constants that should be determined by
comparing with experimental data. As many authors already pointed out
~\cite{EM10,JB150,JV13},
the size of the coupling constant depends on the CP violating model.
Using our study it will be possible, in case of more than one measurements,
to determine the LECs and then individuate the TRV source which generates
the EDM by comparing the values of the calculated LECs and their predicted
sizes.

In future, we plan to use $\chi$EFT to derive the TRV currents which give
contribution at N2LO~\cite{JV11b,JB13}.
This would allow to have a fully consistent
calculation of the EDM of light nuclei up to N2LO. We also plan to study the
$\vec{n}-\vec{p}$ and $\vec{n}-\vec{d}$ spin rotation for an independent
and complementary study of TRV effects respect to EDM. 

\section*{acknowledgments}
Computational resources provided
by the INFN-Pisa Computer Center are gratefully acknowledged.
\appendix
\section{Interaction vertices}
\label{app:vertex}
It is convenient to decompose
the interaction Hamiltonian $H_I$ as follows
\bgroup
\arraycolsep=0.5pt
\begin{eqnarray}
 H_I\!=\!H^{00}\!+\!H^{01}\!+\!H^{10}\!+\!H^{02}\!+\!H^{11}\!+\!H^{20}\!+\!\cdots\ ,
\end{eqnarray}
where $H^{nm}$ has $n$ creation and $m$ annihilation 
operators for the pion. Explicitly,
\begin{eqnarray}
 H^{00} & = & {1\over \Omega}\sum_{\a_1' \a_1\a_2' \a_2} 
 b^{\dag}_{\a_1'}b_{\a_1}  b^{\dag}_{\a_2'} b_{\a_2}
 M^{00}_{\a_1'\a_1\a_2'\a_2}
 \delta_{\bmp_1'+\bmp_2' ,\bmp_1+\bmp_2}\ , \label{eq:m00}\nonumber\\
 \\
 H^{01}& = & \frac{1}{\sqrt{\Omega}} \sum_{\a' \a}\sum_{\bmq \, a} 
   b^{\dag}_{\a'}b_{\a}a_{\bmq\, a}  M^{01}_{\a'\a,\bmq\, a}
   \delta_{\bmq+\bmp,\bmp'}\ ,\label{eq:m01}\\
 H^{10}& = & \frac{1}{\sqrt{\Omega}} \sum_{\a' \a}\sum_{\bmq\, a} 
   b^{\dag}_{\a'}b_{\a}a^\dag_{\bmq\, a}  M^{10}_{\a'\a,\bmq\, a}
   \delta_{\bmq+\bmp',\bmp}\ ,\label{eq:m10}\\
 H^{02} &= & \frac{1}{\Omega}\sum_{\a' \a}\sum_{\bmq' a'\,\bmq\, a}
  b_{\a'}^{\dag}b_{\a}a_{\bmq' a'}a_{\bmq\, a}
  M^{02}_{\a'\a,\bmq'  a'\, \bmq \, a}
 \delta_{\bmq+\bmq'+\bmp,\bmp'}\ ,\label{eq:m02}\nonumber\\
 \\
 H^{11} &= & \frac{1}{\Omega}\sum_{\a' \a}\sum_{\bmq' a'\, \bmq\, a}
  b_{\a'}^{\dag}b_{\a}a^\dag_{\bmq' a'}a_{\bmq\,a }
  M^{11}_{\a'\a,\bmq'a'\bmq \, a}
  \delta_{\bmq+\bmp,\bmq'+\bmp'}\ ,\label{eq:m11}\nonumber\\
  \\
 H^{20} &= & \frac{1}{\Omega}\sum_{\a' \a}\sum_{\bmq'a' \,\bmq\,a}
  b_{\a'}^{\dag}b_{\a}a^\dag_{\bmq' a'}a^\dag_{\bmq \, a}
  M^{20}_{\a'\a,\bmq' a'\, \bmq \,a}
 \delta_{\bmp,\bmq+\bmq'+\bmp'}\ ,\label{eq:m20}\nonumber\\
\end{eqnarray}
\egroup
etc. Here $\alpha_j\equiv \bmp_j,s_j,t_j$ denotes the
momentum, spin projection, isospin projection of nucleon $j$ with energy
$E_j=\sqrt{p_j^2+M^2}$, $\bmq$ and $a$ denote the momentum and isospin
projection of a pion with energy $\omega_q=\sqrt{\bmq^2+m_\pi^2}$, and
$M^{nm}$ are the vertex functions listed below.
The various momenta are discretized by assuming periodic boundary
conditions in a box of volume $\Omega$.  We note that in the expansion
of the nucleon field $\psi$ we have only retained the nucleon degrees of freedom,
since anti-nucleon contributions do not enter the TRV $NN$ and $NNN$
potential at the order
of interest here.  We note also that in general the creation and annihilation
operators are not normal-ordered.  After normal-ordering them,
tadpole-type contributions result, which are relevant only for renormalization,
therefore we discard them hereafter.

The vertex functions $M^{nm}$ involve products of Dirac 4-spinors, which
are expanded non-relativistically in powers
of momenta, and terms up to order $Q^3$ are retained. Useful formulas 
are reported in Appendix F of Ref.~\cite{viviani14}.
\begin{center}
\begin{figure}[h]
  \includegraphics[scale=.4]{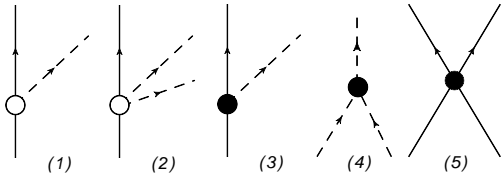}
   \caption{   \label{fig:vertex}
     Vertices entering the TRV potential at N2LO. The
     solid (dash) lines represent nucleons (pions). The open (solid) symbols
     denote PC (TRV) vertices.}
\end{figure}
\end{center}
The interaction vertices needed for the construction of the TRV potential, without
considering renormalization contributions,
are summarized in Fig.~\ref{fig:vertex}.  Note that in the power counting
of these vertices below, we do not include the $1/\sqrt{\omega_k}$ normalization
factors in the pion fields.  We obtain:
\begin{enumerate}
\item $\pi NN$ vertices. The PC interaction term is derived in Appendix F of
  Ref.~\cite{viviani14}. For completeness, here we report the final formula for the
  PC vertex up to order $Q^3$ that reads,
  \begin{eqnarray}
 {}^{PC}M^{\pi NN,01}_{\alpha' \alpha,
     \bmq\,a}  &=&  {g_A \over 2 f_\pi}
 {\tau_a\over \sqrt{2\omega_q}}\Bigl[ i\, \bmq\cdot\bmsi-{i\over
     M}\omega_q\;\bmK\cdot\bmsi \nonumber \\
   &+& {i\over  4M^2}\Bigl(2\bmK\cdot\bmq\;\bmK\cdot\bmsi-
      2K^2\;\bmq\cdot\bmsi\nonumber\\
   && \qquad -{1\over 2}\bmk\cdot\bmsi\; \bmq\cdot\bmk\Bigr)\Bigr]
    \nonumber\\
     &+& {m_\pi^2 \over f_\pi}(2d_{16}-d_{18})
    {\tau_a\over \sqrt{2\omega_q}}\; i\bmq\cdot\bmsi\ ,\label{eq:MpiNN01b}\\
   {}^{PC}M^{\pi NN, 10}_{\alpha'\alpha,
     \bmq\,a}   &= &  -{}^{PC}M^{\pi NN,01}_{\alpha' \alpha,\bmq\,a}  .
  \label{eq:MpiNN10b}
  \end{eqnarray}
  In diagrams, these PC vertex functions are represented as open circles.
  The TRV $\pi NN$ vertices are due to interaction terms proportional to LECs
  $g_0$, $g_1$ and $g_2$ which corresponds to isoscalar, isovector and isotensor
  interaction, plus terms which derive from ${\cal L}_{\text{TRV}}^{\pi N\, (1)}$
  and ${\cal L}_{\text{TRV}}^{\pi N\, (2)}$ given in Eqs.~(\ref{eq:ltrv1}) and
  ~(\ref{eq:ltrv2}).
  They read (up to order $Q^2$),
  \begin{align}
    {}^{TRV}&M^{\pi NN,01}_{\alpha' \alpha,
      \bmq\,a}=-\frac{(g_0 \tau_a+g_1\delta_{a,3}+g_2\tau_3\delta_{a,3})}
    {\sqrt{2\omega_q}}\nonumber\\
    &\qquad\times\Bigl[ 1-{1\over  4M^2}\Bigl(
      2K^2+i\ \bmk\times\bmK\Bigr)\Bigr]\nonumber\\
    &\qquad-ig^{(1)}_V\frac{\epsilon_{ab3}\tau_b}{\fp\sqrt{2\omega_q}}\big[\omega_q
      -{1\over 2M}\bmK\cdot\bmq\big]\nonumber\\
    &\qquad+\frac{\mp^2}{\fp^2\sqrt{2\omega_q}}\big(g^{(2)}_{S1}\tau_a
    +g^{(2)}_{V1}\delta_{a,3}+g^{(2)}_{T1}\tau_3\delta_{a,3}\big)\nonumber\\
    &\qquad-\frac{\omega_q^2}{\fp^2\sqrt{2\omega_q}}\big(g^{(2)}_{S2}\tau_a
    +g^{(2)}_{V2}\delta_{a,3}+g^{(2)}_{T2}\tau_3\delta_{a,3}\big)
    \ ,\nonumber\\\label{eq:MpiNN01a}
  \end{align}
  \begin{align}
   {}^{TRV}&M^{\pi NN, 10}_{\alpha'\alpha,
     \bmq\,a}=-\frac{(g_0 \tau_a+g_1\delta_{a,3}+g_2\tau_3\delta_{a,3})}
    {\sqrt{2\omega_q}}\nonumber\\
    &\qquad\times\Bigl[ 1-{1\over  4M^2}\Bigl(
      2K^2+i\ \bmk\times\bmK\Bigr)\Bigr]\nonumber\\
    &\qquad+ig^{(1)}_V\frac{\epsilon_{ab3}\tau_b}{\fp\sqrt{2\omega_q}}\big[\omega_q
      -{1\over 2M}\bmK\cdot\bmq\big]\nonumber\\
    &\qquad+\frac{\mp^2}{\fp^2\sqrt{2\omega_q}}\big(g^{(2)}_{S1}\tau_a
    +g^{(2)}_{V1}\delta_{a,3}+g^{(2)}_{T1}\tau_3\delta_{a,3}\big)\nonumber\\
    &\qquad-\frac{\omega_q^2}{\fp^2\sqrt{2\omega_q}}\big(g^{(2)}_{S2}\tau_a
    +g^{(2)}_{V2}\delta_{a,3}+g^{(2)}_{T2}\tau_3\delta_{a,3}\big)\ ,\nonumber\\
   \label{eq:MpiNN10a}
  \end{align}
\item $\pi\pi NN$ vertices. The PC interaction is needed up to NLO in the
  NR expansion at the order we are interested. The corresponding vertex functions
  will receive contribution from the Weinberg-Tomozawa term and from 
  ${\cal L}^{(2)}_{N\pi}$. The vertex functions read,
  \begin{align}
   {}^{PC}M&^{\pi\pi NN,02}_{\alpha' \alpha,\bmq' a'\,\bmq\, a}=
    {i\over 8f_\pi^2}  {\epsilon_{aa'b}\tau_b
      \over\sqrt{2\omega_q}\sqrt{2\omega_{q'}}}\Big[(\omega_q-\omega_{q'})
      \nonumber\\
      &-\frac{2\bmK\cdot(\bmq-\bmq')-i(\bmk\times\bmsi)\cdot(\bmq-\bmq')}
          {2M}\Big]\nonumber\\
    &+\frac{\delta_{ij}}{\fp^2}\frac{c_1(2m_\pi^2)+c_2\omega_q\omega_{q'}+c_3
      (\omega_q\omega_{q'}-\bmq\cdot\bmq')}{\sqrt{2\omega_q}\sqrt{2\omega_{q'}}}
    \nonumber\\
    &-\frac{c_4}{2\fp^2}{\epsilon_{aa'b}\tau_b
    \over\sqrt{2\omega_q}\sqrt{2\omega_{q'}}}(\bmq\times\bmq')\cdot\bmsi
    \ , \\ 
   {}^{PC}M&^{\pi\pi NN, 20}_{\alpha' \alpha,\bmq' a'\,\bmq\, a}=
    {i\over 8f_\pi^2}  {\epsilon_{aa'b}\tau_b
      \over\sqrt{2\omega_q}\sqrt{2\omega_{q'}}}\Big[(\omega_{q'}-\omega_{q})
      \nonumber\\
      &-\frac{2\bmK\cdot(\bmq'-\bmq)-i(\bmk\times\bmsi)\cdot(\bmq'-\bmq)}
          {2M}\Big]\nonumber\\
    &+\frac{\delta_{ij}}{\fp^2}\frac{c_1(2m_\pi^2)+c_2\omega_q\omega_{q'}+c_3
      (\omega_q\omega_{q'}-\bmq\cdot\bmq')}{\sqrt{2\omega_q}\sqrt{2\omega_{q'}}}
    \nonumber\\
    &-\frac{c_4}{2\fp^2}{\epsilon_{aa'b}\tau_b
    \over\sqrt{2\omega_q}\sqrt{2\omega_{q'}}}(\bmq\times\bmq')\cdot\bmsi
    \ .
  \end{align}
  The ${}^{PC}M^{\pi\pi NN,11}_{\alpha' \alpha,\bmq' a'\,\bmq\, a}$ vertex
  is not needed
  for the evaluation of the time-ordered diagrams.
\item $3\pi$ vertices. The TRV Lagrangian has a three-pion interaction term
  shown in diagram (4) of Fig.~\ref{fig:vertex}.
  The vertices, neglecting the tadpole terms,
  are given by $H_I^{3\pi}=H_I^{3\pi,03}+H_I^{3\pi,12}
  +H_I^{3\pi,21}+H_I^{3\pi,30}$,
  where
  \begin{align}
    H_I^{3\pi,03} &={1\over\Omega^{3/2}}
    \sum_{\substack{\bmq a\,\bmq'a'\\ \bmp b}} a_{\bmq\,a}a_{\bmq'\,a'}a_{\bmp\,b}\;
    M^{3\pi,03}_{\bmq a\,\bmq'a'\,\bmp b}\,\delta_{\bmq+\bmq'+\bmp,0}\ ,\nonumber\\
    \\
    H_I^{3\pi,12} &= {1\over\Omega^{3/2}}
    \sum_{\substack{\bmq a\,\bmq'a'\\ \bmp b}} a_{\bmq\,a}a_{\bmq'\,a'}a_{\bmp\,b}^\dagger\;
    M^{3\pi,12}_{\bmq a\,\bmq'a'\, \bmp b}\,\delta_{\bmq+\bmq',\bmp}\ ,\nonumber\\
    \\
    H_I^{3\pi,21} &= {1\over\Omega^{3/2}}
    \sum_{\substack{\bmq a\,\bmq'a'\\\bmp b}} a_{\bmq\,a}a_{\bmq'\,a'}^\dagger a_{\bmp\,b}^\dagger\;
    M^{3\pi,21}_{\bmq a\,\bmq'a'\,\bmp b}\,\delta_{\bmq,\bmq'+\bmp}\ ,\nonumber\\
    \\
    H_I^{3\pi,30} & ={1\over\Omega^{3/2}}
    \sum_{\substack{\bmq a\,\bmq'a'\\ \bmp b}} a_{\bmq\,a}^\dagger a_{\bmq'\,a'}^\dagger a_{\bmp\,b}^\dagger\;
    M^{3\pi,30}_{\bmq a\,\bmq'a'\,\bmp b}\,\delta_{0,\bmq+\bmq'+\bmp}\ ,\nonumber\\
  \end{align}
  with,
  \begin{align}
    {}^{TRV}M^{3\pi,03}_{\bmq a\,\bmq'a'\,\bmp b}&=-\frac{\Delta M}
    {3\sqrt{2\omega_q\,2\omega_{q'}\,2\omega_p}}\nonumber\\
    &\times\bigl(\delta_{a,a'}\delta_{b,3}+
    \delta_{a,b}\delta_{a',3}+\delta_{a',b}\delta_{a,3}\bigr)\ ,\nonumber\\
    \\
         {}^{TRV}M^{3\pi,12}_{\bmq a\,\bmq'a'\,\bmp b}&=
          3\,{}^{TRV}M^{3\pi,03}_{\bmq a\,\bmq'a'\,\bmp b}\ ,\\
         {}^{TRV}M^{3\pi,21}_{\bmq a\,\bmq'a'\,\bmp b}&=
          3\,{}^{TRV}M^{3\pi,03}_{\bmq a\,\bmq'a'\,\bmp b}\ ,\\
         {}^{TRV}M^{3\pi,30}_{\bmq a\,\bmq'a'\,\bmp b}&=
          {}^{TRV}M^{3\pi,03}_{\bmq a\,\bmq'a'\,\bmp b}\ .
  \end{align}
\item $4N$ contact interaction. The EFT Hamiltonian includes also the term
  given in Eq.~(\ref{eq:m00}) derived from a contact Lagrangian.
  We only need its TRV part of order $Q$, which includes
  five independent interaction terms.
  With a suitable choice of the LECs, 
  the vertex function ${}^{TRV}M^{00}$ can be written as
  \begin{align}
    {}^{TRV}&M^{00}_{\a_1'\a_1\a_2'\a_2}=  
    {1 \over 2\Lambda_\chi^2 f_\pi} \Bigl[ 
      C_1 i \bmk_1\cdot (\sone-\stwo) \nonumber\\
      &\qquad+ C_2\,\tone\cdot\ttwo\, i\bmk_1\cdot(\sone-\stwo)\nonumber\\
      &\qquad+  {C_3\over 2}\, \bigl(\left(\tau_{1z}+\tau_{2z}\right)
      \, i \bmk_1\cdot\left(\sone-\stwo\right)\nonumber\\
      &\qquad\qquad-\left(\tau_{1z}-\tau_{2z}\right)\,
      i\bmk_1\cdot\left(\sone+\stwo\right)\bigr)\nonumber\\
      &\qquad+  {C_4\over 2}\, \bigl(\left(\tau_{1z}+\tau_{2z}\right)
      \, i \bmk_1\cdot\left(\sone-\stwo\right)\nonumber\\
      &\qquad\qquad+\left(\tau_{1z}-\tau_{2z}\right)\,
      i\bmk_1\cdot\left(\sone+\stwo\right)\bigr)\nonumber\\
       &\qquad+C_5\, \left(3\tau_{1z}\tau_{2z}-\tone\cdot\ttwo\right) i\bmk_1\cdot
      \left(\sone-\stwo\right)
      \Bigr]\ , \label{eq:m00trv}\nonumber\\
  \end{align}
  where $\bmk_1=\bmp_1'-\bmp_1=-\bmp_2'+\bmp_2$.
\end{enumerate}

\section{The TRV $NN$ potential}\label{app:pot}
In this section we discuss the derivation of the TRV $NN$ and $NNN$ potential,
providing explicit expressions of the diagrams given in Fig.~\ref{fig:diagNN}.
The power counting is as follows:
(i) a PC (TRV) $\pi NN$ vertex is of order $Q$ ($Q^{0}$);
(ii) a PC $\pi\pi NN$ vertex is of order $Q^1$;
(iii) a TRV $3\pi$ vertex is of order $Q^{-3}$;
(iv) a PC (TRV) $NN$ contact vertex is of order $Q^0$ ($Q$);
(v) an energy denominator without (with one or more)
pions is of order $Q^{-2}$ ($Q^{-1}$);
(vi) factors $Q^{-1}$ and $Q^{3}$ are associated with,
respectively, each pion line and each loop
integration.
The momenta are defined as given in Eq.~(\ref{eq:notjb1}), and in what follows
use is made of the fact that $\bmk\cdot\bmK$ vanishes in the c.m. frame. It is
useful to define the isospin operator as,
\begin{eqnarray}
  \TO &=&\tone\cdot\ttwo\ ,\\
  \ttp&=&(\tau_{1z}+\tau_{2z})\ ,\\
  \ttm&=&(\tau_{1z}-\tau_{2z})\ ,\\
  \tten&=&(3\tau_{1z}\tau_{2z}-\tone\cdot\ttwo)\ .
\end{eqnarray}

The TRV $NN$ potential is derived from the amplitudes in Fig.~\ref{fig:diagNN}
via Eqs.~(\ref{eq:vtrvml})--(\ref{eq:vtrv1}).
Up to order $Q$ included, we obtain for the OPE
component in panel (a) of Fig.~\ref{fig:diagNN} :
\begin{equation}
 V({\rm a}) =
 V^{(-1)}({\rm NR})+ V^{(1)}({\rm RC}) + V^{(1)}({\rm LEC})\ , \label{eq:ope2}
\end{equation}
where,
 \begin{align}
 V^{(-1)}&({\rm NR})=\frac{\ga g_0^*}{2\fp}\TO\frac{i\bmk\cdot
    (\sone-\stwo)}{\omk^2} \nonumber\\
 &+\frac{g_Ag_1}{4f_\pi}\Big[\ttp
    \frac{i\bmk\cdot(\sone-\stwo)}{\omega_k^2}+
    \ttm\frac{i \bmk\cdot(\sone+\stwo)}{\omega_k^2}\Big]\nonumber\\
 &+\frac{\ga g_2}{6\fp}\tten\frac{i\bmk\cdot
    (\sone-\stwo)}{\omk^2}  
 \!\!\ ,\label{eq:opeloapp}
 \end{align}
 \begin{align}
 \vtrv^{(1)}&{(\rm RC)}=\frac{\ga g_0^*}{8\fp M^2}
  \TO\ \frac{1}{\omk^2}\nonumber\\
  &\times\Big[-\frac{i}{2} \left(8K^2+k^2\right)\bmk\cdot(\sone-\stwo)\nonumber\\
   &+\bmk\cdot\sone(\bmk\times\bmK)\cdot\stwo
    -\bmk\cdot\stwo(\bmk\times\bmK)\cdot\sone\Big]\nonumber\\
  &+\frac{\ga g_1}{16\fp M^2}\ \frac{1}{\omk^2}\nonumber\\
  &\times\Big\{\ttm
  \Big[-\frac{i}{2} \left(8K^2+k^2\right)\bmk\cdot(\sone+\stwo)\nonumber\\
   &+\bmk\cdot\sone(\bmk\times\bmK)\cdot\stwo
    +\bmk\cdot\stwo(\bmk\times\bmK)\cdot\sone\Big]\nonumber\\
  &+\ttp
  \Big[-\frac{i}{2} \left(8K^2+k^2\right)\bmk\cdot(\sone-\stwo)\nonumber\\
   &+\bmk\cdot\sone(\bmk\times\bmK)\cdot\stwo
    -\bmk\cdot\stwo(\bmk\times\bmK)\cdot\sone\Big]\Big\}\nonumber\\
  &+\frac{\ga g_2}{24\fp M^2}\tten
  \ \frac{1}{\omk^2}
  \Big[-\frac{i}{2} \left(8K^2+k^2\right)\bmk\cdot(\sone-\stwo)\nonumber\\
   &+\bmk\cdot\sone(\bmk\times\bmK)\cdot\stwo
    -\bmk\cdot\stwo(\bmk\times\bmK)\cdot\sone\Big]\label{eq:rcopeapp}\ ,\\
  \vtrv^{(1)}&({\rm LEC})=\vtrv^{(-1)}({\rm NR})
  \frac{2m^2_\pi}{\ga}(2d_{16}-d_{18})\nonumber\\
  &+\frac{\ga}{2\fp^3}\TO i\bmk\cdot(\sone-\stwo)
  \Big[g^{(2)}_{S2}-g^{(2)}_{S1}\frac{\mp^2}{\omk^2}\Big]\nonumber\\
 &+\frac{g_A}{4f_\pi^3}\Big[\ttp
    i\bmk\cdot(\sone-\stwo)+
    \ttm i \bmk\cdot(\sone+\stwo)\Big]\nonumber\\
 &\qquad \times\Big[g^{(2)}_{V2}-g^{(2)}_{V1}\frac{\mp^2}{\omk^2}\Big]\nonumber\\
 &+\frac{\ga}{6\fp^3}\tten i\bmk\cdot(\sone-\stwo)
  \Big[g^{(2)}_{T2}-g^{(2)}_{T1}\frac{\mp^2}{\omk^2}\Big]\nonumber\\
  &+\frac{g^{(1)}_Vg_A}{2\fp^2M}(\tone\times\ttwo)_z\,\bmK\cdot(\sone+\stwo)\,
  \label{eq:lec},
\end{align}
 where $g_0^*=g_0+g_2/3$.
 The contribution given in the first line of Eq.~(\ref{eq:lec})
 renormalizes 
 the coupling constants $g_0^*$, $g_1$ and $g_2$ in the OPE
 term. The terms of Eq.~(\ref{eq:lec}) which are multiplied by the factor
 $\mp^2/\omk^2$ are also reabsorbed in the constant $g_0^*$, $g_1$ and $g_2$ while
 all the other terms are reabsorbed in  the LECs $C_2$, $C_4$ and $C_5$ in
 Eq.~(\ref{eq:ct}). Regarding the last term in Eq.~(\ref{eq:lec}),
 it is possible to use a Fierz transformation
 obtaining a combination of the operators which multiplies the LECs $C_3$
 and $C_4$. Therefore all $\vtrv^{(1)}(\text{LEC})$  can be reabsorbed in the
 OPE and contact potentials.
 The factor $k^2=\omega_k^2-m_\pi^2$ in
 the isoscalar, isovector and isotensor component of
 $V^{(1)}({\rm RC})$ leads to a piece that can be reabsorbed in
 the contact term proportional to
 $C_2$, $C_4$ and $C_5$ in Eq.~(\ref{eq:ct})
 and a piece proportional to $m_\pi^2$ that simply
 renormalizes the LECs $g_0^*$, $g_1$ and $g_2$. 

 The component of the TRV potential coming from the contact
 terms in panel (b) of Fig.~\ref{fig:diagNN} derives directly from the vertex
 function ${}^{TRV}M^{00}$ given in Eq.~(\ref{eq:m00trv}).
 The final expression has
 already been given in Eq.~(\ref{eq:ct}).
 The diagrams reported in panels (c) and (d) contain a combination of a contact
 interaction with the exchange of a pion. However, it can be
 shown that their contribution is at least of order $Q^3$ .
 
 Next we consider the TPE components in panels (e)-(h). The contribution from
 diagrams (e) reads,
 \begin{eqnarray}
   V^{(1)}(\rm e)&=&-\frac{\ga g_0^*}{4\fp^3}\TO \, i\bmk\cdot
   (\sone-\stwo)\int_\bmq\frac{1}{\omp\omm(\omp+\omm)} \nonumber\\
   &&+\frac{\ga g_2}{12\fp^3}\tten \, i\bmk\cdot
   (\sone-\stwo)\int_\bmq\frac{1}{\omp\omm(\omp+\omm)}\ ,
   \nonumber\\
 \end{eqnarray}
 where $\omega_\pm=\sqrt{(\bmk\pm \bmq)^2 + 4 \, m_\pi^2}$ and
 $\int_\bmq=\int \frac{d\bmq}{(2\pi)^3}$. The
 isovector component of the OPE vertex vanishes since the integrand
 is proportional to $\bmq$,
 therefore there is no isovector component from panel (e).
 Dimensional regularization allows one to obtain the finite part as,
 \begin{eqnarray}
  \overline{V}^{(1)}(\rm e)&=&\frac{\ga g_0^*}{\fp\Lx^2}\TO i\bmk\cdot
  (\sone-\stwo) L(k)\nonumber\\
  &&-\frac{\ga g_2}{3\fp\Lx^2}\tten i\bmk\cdot
  (\sone-\stwo) L(k)\ ,\label{eq:pote}
 \end{eqnarray}
 where $\Lambda_\chi=4\pi f_\pi$ and the loop function $L(k)$ is defined as
 \begin{equation}
   L(k)= {1\over 2} {s\over k} \ln\left({s+k\over s-k}\right)\ ,\quad
   s=\sqrt{k^2+4\, m^2_\pi}\ .\label{eq:sL}
 \end{equation}
 The singular part is given by,
 \begin{eqnarray}
   V^{(1)}_\infty({\rm e}) &=&\frac{\ga g_0^*}{2\fp\Lx^2}\TO i\bmk\cdot
  (\sone-\stwo)(\deps-2)\nonumber\\
   &&-\frac{\ga g_2}{6\fp\Lx^2}\tten i\bmk\cdot
  (\sone-\stwo)(\deps-2)\ ,
 \end{eqnarray}
 where
\begin{equation}
  \deps =  -{2\over \epsilon}+\gamma-\ln
  \pi +\ln\left({m^2_\pi\over\mu^2}\right)
      \ ,
\end{equation}
$\epsilon=3-d$, $d$ being the number of dimensions ($d\rightarrow 
3$), and $\mu$ is a renormalization scale.  This singular contribution
is absorbed in the $\vtrv^{({\rm CT})}$ term proportional
to $C_2$ for the isoscalar part and to $C_5$ for the isotensor part.

The contributions from panels (f)-(h) in Fig.~\ref{fig:diagNN}
are collectively denoted
as ``box'' below, and the non-iterative pieces in
reducible diagrams of type (h) are identified via Eq.~(\ref{eq:vtrv1}). From the
panel (f) we obtain,
\begin{align}
  V^{(1)}&({\rm f})=\Big\{-\frac{\ga^3 g_0^*}{16\fp^3}(3+2\ \TO)i\bmk\cdot
  (\sone-\stwo)\nonumber\\
  &+\frac{\ga^3 g_2}{48\fp^3}\tten \ i\bmk\cdot
  (\sone-\stwo)\nonumber\\
  &-\frac{\ga^3 g_1}{32\fp^3}
  \Big[\ttp \ i\bmk\cdot(\bmsi_1-\bmsi_2)+
    \ttm \ i\bmk\cdot(\bmsi_1+\bmsi_2)\Big]\Big\}\nonumber\\
  &\times\int_\bmq\frac{\omp^2+\omp\omm+\omm^2}{\omp^3\omm^3(\omp+\omm)}
  (k^2-q^2)\ ,\label{eq:panelf}
\end{align}
while the contribution of panel (g) results,
\begin{align}
  V^{(1)}&({\rm g})=\Big\{\frac{\ga^3 g_0^*}{16\fp^3}(3-2\ \TO)i\bmk\cdot
  (\sone-\stwo)\nonumber\\
  &+\frac{\ga^3 g_2}{48\fp^3}\tten \ i\bmk\cdot
  (\sone-\stwo)\nonumber\\
  &+\frac{\ga^3 g_1}{32\fp^3}
  \Big[\ttp \ i\bmk\cdot(\bmsi_1-\bmsi_2)+
    \ttm \ i\bmk\cdot(\bmsi_1+\bmsi_2)\Big]\Big\}\nonumber\\
  &\times\int_\bmq\frac{\omp^2+\omp\omm+\omm^2}{\omp^3\omm^3(\omp+\omm)}
  (k^2-q^2)\ .\label{eq:panelg}
\end{align}
The complete ``box'' contribution is given by the sum of $V({\rm f})$ and $V({\rm g})$
and it reads,
\begin{align}
  V^{(1)}({\rm box})&=\Big\{-\frac{\ga^3 g_0^*}{4\fp^3}\ \TO \ i\bmk\cdot
  (\sone-\stwo)\nonumber\\
  &+\frac{\ga^3 g_2}{12\fp^3}\ \tten \ i\bmk\cdot
  (\sone-\stwo)\Big\}\nonumber\\
  &\times\int_\bmq\frac{\omp^2+\omp\omm+\omm^2}{\omp^3\omm^3(\omp+\omm)}
  (k^2-q^2)\ ,\label{eq:panelbox}
\end{align}
where all the isovector terms cancel out.
After dimensional regularization,
the finite part reads,
\begin{align}
  \overline{V}^{(1)}({\rm box})&=\frac{\ga^3 g_0^*}{\fp\Lx^2}\ \TO \ i\bmk\cdot
  (\sone-\stwo) [H(k)-3\, L(k)]\nonumber\\
  &-\frac{\ga^3 g_2}{3\fp\Lx^2}\ \tten \ i\bmk\cdot
  (\sone-\stwo) [H(k)-3\, L(k)]\ ,\label{eq:box}\nonumber\\
\end{align}
where
\begin{equation}
   H(k)= {4\, m^2_\pi\over s^2} L(k)\ ,\label{eq:H}
\end{equation}
while the singular part is given by,
\begin{align}
  V^{(1)}_{\infty}&({\rm box})=-\frac{\ga^3 g_0^*}{\fp\Lx^2}\TO\
  i\bmk\cdot(\sone-\stwo)\Big(\frac{3}{2}\deps-1\Big)\nonumber\\
  &+\frac{\ga^3 g_2}{3\fp\Lx^2}\tten\
  i\bmk\cdot(\sone-\stwo)\Big(\frac{3}{2}\deps-1\Big)\ .
\end{align}
The latter is absorbed in the $\vtrv^{({\rm CT})}$ term proportional
to $C_2$ for the isoscalar part and to $C_5$ for the isotensor part.

Now we consider the contributions that come from the panels (i) and (l) of Fig.
~\ref{fig:diagNN}.
At NLO the contributions of the panel (l) cancel out due to the isospin
structure of the vertices. The contribution of diagrams (i) result,
\begin{align}
  V^{(0)}({\rm i})&=-\frac{5\ga^3\Delta_3M}{32\fp^3}\frac{1}{\omk^2}
  \Big[\ttp\ i\bmk\cdot(\bmsi_1-\bmsi_2)\nonumber\\
  &+\ttm \ i \bmk\cdot(\bmsi_1+\bmsi_2)\Big] 
  \int_\bmq \frac{k^2-q^2}{\omp^2\omm^2}\ .\label{eq:3piapp}
\end{align}
Using dimensional regularization we obtain,
\begin{align}
  \overline{V}^{(0)}&({\rm i})=-\frac{5\ga^3\Delta_3 M}{4\fp\Lx^2}
  \frac{\pi}{\omega_k^2}
  \big(\ttm\ i \bmk\cdot(\sone-\stwo)\nonumber\\
  &+\ttp\ i \bmk\cdot(\sone-\stwo)\big)
  \Big[\Big(1-\frac{2\mp^2}{s^2}\Big)s^2A(k)+\mp\Big]\ ,\nonumber\\
\end{align}
where,
\begin{eqnarray}
  A(k)=\frac{1}{2k}\arctan\Big({\frac{k}{2\mp}}\Big)\ .\label{eq:A}
\end{eqnarray}
To be noticed that the use of dimensional regularization does not give the
divergent part of the integral in Eq.~(\ref{eq:3piapp}). This is due to the fact
that  the dimensional regularization cannot deal with linear divergences.
To explicit the linear divergence we use a simple regularization of
Eq.~(\ref{eq:3piapp}), namely we integrate over $q$ up to a (large) value $\Lambda_R$.
The result for the non divergent
part is equal to the one reported in Eq.~(\ref{eq:3piapp}) while the divergent
part reads,
\begin{align}
  V^{(0)}_{\infty}&({\rm i})=-\frac{5\ga^3\Delta_3 M}{4\fp\Lx^2}
  \frac{1}{\omega_k^2}
  \big(\ttm\ i \bmk\cdot(\sone-\stwo)\nonumber\\
  &+\ttp\ i \bmk\cdot(\sone-\stwo)\big)
  \Big[\Lambda_R+{\cal O}\big(\frac{k^2}{\Lambda_R}\big)\Big]\ ,
\end{align}
where spurious contributions of order $Q^2/\Lambda_R$ or more appear but they
can be neglected for $\Lambda_R\rightarrow\infty$. The divergent part can be
reabsorbed in the $\vtrv^{({\rm CT})}$ term proportional to $C_4$.

At N2LO the contribution of panel (i) comes both from the second order
in the pion propagator (PP) and in the pion-nucleon vertex (PNV).
For the former we obtain,
\begin{align}
  &V^{(1)}({\rm i-PP})=-\frac{5\ga^3\Delta_3}{128\fp^3}\frac{1}{\omk^2}
  \Big[\ttp\ 
    i\bmk\cdot(\bmsi_1-\bmsi_2)\nonumber\\
    &\quad+\ttm \ i \bmk\cdot(\bmsi_1+\bmsi_2)\Big]
  \int_\bmq \frac{\omp^2+\omp\omm+\omm^2}{\omp^3\omm^3(\omp+\omm)}
  \big(k^2-q^2\big)^2\nonumber\\
  &\quad+\frac{5\ga^3\Delta_3}{8\fp^3}\frac{1}{\omk^2}
  \int_\bmq \frac{\omp^2+\omp\omm+\omm^2}{\omp^3\omm^3(\omp+\omm)}\nonumber\\
  &\qquad\qquad\times
  \big[\bmk\cdot\sone(\bmq\times\bmk)\cdot\stwo(\bmq\cdot\bmK)\tau_{1z}
    \nonumber\\
    &\qquad\qquad
    -\bmk\cdot\stwo(\bmq\times\bmk)\cdot\sone(\bmq\cdot\bmK)\tau_{2z}\big]
\end{align}
while for the latter,
\begin{align}
  &V^{(1)}({\rm i-PNV})=\frac{5\ga^3\Delta_3}{64\fp^3}\frac{1}{\omk^2}
  \Big[\ttp\ 
    i\bmk\cdot(\bmsi_1-\bmsi_2)\nonumber \\
    &\qquad+\ttm \ i \bmk\cdot(\bmsi_1+\bmsi_2)\Big]
  \int_\bmq \frac{k^2+q^2}{\omp\omm(\omp+\omm)}\nonumber\\
  &\qquad+\frac{5\ga^3\Delta_3}{16\fp^3}\frac{1}{\omk^2}\nonumber\\
  &\qquad\times\Big[\big(\bmk\cdot\sone(\bmk\times\bmK)\cdot\stwo
    +\bmk\cdot\stwo(\bmk\times\bmK)\cdot\sone\big)\ttm\nonumber\\
    &\qquad+\big(\bmk\cdot\sone(\bmk\times\bmK)\cdot\stwo
   -\bmk\cdot\stwo(\bmk\times\bmK)\cdot\sone\big)\ttp\Big]\nonumber\\
  &\qquad\times\int_\bmq \frac{1}{\omp\omm(\omp+\omm)}\ .
\end{align}
Also in this case we use the dimensional regularization that permits us to
write the finite contribution as,
\begin{align}
  \overline{V}&^{(1)}({\rm i})=-\frac{5\ga^3\Delta_3}{16\fp\Lx^2}
  (\ttm\ i\bmk\cdot(\bmsi_1+\bmsi_2) \nonumber\\
  &+\ttp\ i\bmk\cdot(\bmsi_1-\bmsi_2))\nonumber\\
  &\times\Big(\frac{25}{6}L(k)-\frac{7}{2}\frac{\mp^2L(k)}{\omega_k^2}
  +2\frac{\mp^2H(k)}{\omega_k^2}\Big)
  -\frac{25\ga^3\Delta_3}{12\fp\Lx^2}\frac{1}
  {\omega_k^2}\nonumber\\
  &\times\Big[\ttm \ \big((\bmk\times\bmK)\cdot\stwo\bmk\cdot\sone
    +(\bmk\times\bmK)\cdot\sone\bmk\cdot\stwo\big)\nonumber\\
    &+\ttp \ \big((\bmk\times\bmK)\cdot\stwo\bmk\cdot\sone
    -(\bmk\times\bmK)\cdot\sone\bmk\cdot\stwo\big)\Big]\ ,
\end{align}
while the divergent part reads,
\begin{align}
  V^{(1)}_\infty&({\rm i})=
  \frac{5\ga^3\Delta_3}{16\fp\Lx^2}
  (\ttm\ i\bmk\cdot(\bmsi_1+\bmsi_2)\nonumber\\
  &+\ttp\ i\bmk\cdot(\bmsi_1-\bmsi_2))\nonumber\\
  &\times\Big[\frac{\mp^2}
    {\omega_k^2}\Big(\frac{151}{12}\deps-\frac{305}{18}\Big)
    -\frac{25}{12}\deps-\frac{2}{9}\Big]\ .
  \label{eq:divpanli}
\end{align}
The divergences present in Eq.~(\ref{eq:divpanli}) are reabsorbed in the
$\vtrv^{({\rm OPE})}$ term proportional to $g_1$ for the part which multiply
 $\mp^2/\omega_k^2$ and in the
$\vtrv^{({\rm CT})}$ term proportional to $C_4$ for the rest.
All the divergences related to the term where $\bmK$ is present cancel out.

The N2LO contribution of panel (l) in Fig.~\ref{fig:diagNN} is proportional to
the LECs $c_1$, $c_2$ and $c_3$ of the PC sector and it reads,
\begin{align}
  V^{(1)}&({\rm l})=-\frac{5\ga\Delta_3M}{2\fp^3}\frac{1}{\omk^2}
  \Big[\ttp \ i\bmk\cdot(\bmsi_1-\bmsi_2)\nonumber\\
  &+\ttm \ i \bmk\cdot(\bmsi_1+\bmsi_2)\Big]
  \Big(c_1 m_\pi^2\int_\bmq\frac{1}{\omp\omm(\omp+\omm)}\nonumber\\
  &+\frac{c_2+c_3}{8}\int_\bmq\frac{1}{\omp+\omm}
  -\frac{c_3}{8}\int_\bmq\frac{k^2-q^2}{\omp\omm(\omp+\omm)}\Big)\ .
\end{align}
Using dimensional regularization we obtain for the finite part,
\begin{align}
  \overline{V}^{(1)}&({\rm l})=\frac{5\ga\Delta_3M}{2\fp\Lx^2}
  (\ttm\ i\bmk\cdot(\bmsi_1+\bmsi_2)\nonumber\\
  &+\ttp\ i\bmk\cdot(\bmsi_1+\bmsi_2))\nonumber\\
  &\times\Big[4c_1\ \frac{\mp^2L(k)}{\omega_k^2}
       -\frac{c_2}{3}\
       \Big(2L(k)+6\frac{\mp^2}{\omega_k^2}L(k)\Big)\nonumber\\
       &-\frac{c_3}{2}\ 
       \Big(3L(k)+5\frac{\mp^2}{\omega_k^2}L(k)\Big)\Big]\ ,
\end{align}
while the divergent part is given by,
\begin{align}
  V^{(1)}_\infty&(l)=\frac{5\ga\Delta_3M}{2\fp\Lx^2}
  (\ttm\ i\bmk\cdot(\bmsi_1+\bmsi_2)\nonumber\\
  &+\ttp\ i\bmk\cdot(\bmsi_1-\bmsi_2))
  \Big[c_1\frac{1}{\omega_k^2}(-2\deps+4)\nonumber\\
    &-\frac{c_2}{3}\Big(\frac{\mp^2}{\omega_k^2}
    \big(5\deps-\frac{19}{3}\big)
         +\big(\deps-\frac{5}{3}\big)\Big)\nonumber\\
       &-\frac{c_3}{2}\Big(\frac{\mp^2}
         {\omega_k^2}\big(\frac{11}{2}\deps-\frac{25}{3}\big)
         +\big(\frac{3}{2}\deps-\frac{2}{3}\big)\Big)\Big]\ .
\end{align}
As for the panel (i) the divergences which multiply $\mp^2/\omega_k^2$
are reabsorbed in the
$V^{({\rm OPE})}$ term proportional to $g_1$ while all the others in the
$V^{({\rm CT})}$ term proportional to $C_4$. At N2LO panel (l) can receive
contribution also from the second order in the $\pi NN$ and $\pi\pi NN$ vertices
but due to the isospin structure these contributions vanish.

As regarding the $NNN$ TRV potential all the contributions come from diagrams (m)
of Fig.~\ref{fig:diagNN}. The expression we obtain at NLO is given in Eq.~(\ref{eq:NNNpot}).
The N2LO component would come from NLO PC
$\pi NN$ vertex or in the pion propagators. In both cases the different
time-order diagrams cancel out each-other completely.

\section{The potential in configuration space}\label{app:rpot}

\subsection{The $NN$ potential}\label{app:rpotNN}
In this subsection we present the $NN$ potential part in the
configuration space which follows
directly from Eq.~(\ref{eq:vrsp}) and it reads,
\begin{eqnarray}
 V_{TRV}(\bmr,\bmp)&=&
 V^{({\rm OPE})}(\bmr)+V^{({\rm TPE})}(\bmr)+V^{(3\pi,0)}(\bmr)
 \nonumber\\
 &+&V^{(3\pi,1)}(\bmr)+
 V^{({\rm CT})}(\bmr)+V^{({\rm RC})}(\bmr,\bmp)
 \nonumber\\
 &+&V^{(3\pi,{\rm RC})}(\bmr,\bmp)\ ,
 \label{eq:trvnnr}
\end{eqnarray}
where $\bmp=-i\bmna$ is the relative momentum operator. It is convenient to
define the operators,
\begin{eqnarray}
  \ssp&=&(\sone+\stwo)\cdot\bmvr\ ,\\
  \ssm&=&(\sone-\stwo)\cdot\bmvr\ ,\\
  \slp&=&-i(\sone+\stwo)\cdot(\bmvr\times\hat{\bm L})\ ,\\
  \slm&=&-i(\sone-\stwo)\cdot(\bmvr\times\hat{\bm L})\ ,\\
  \srx&=&(\sone\times\stwo)\cdot\bmvr\ ,\\
  \srlp&=&\sone\cdot\bmvr\, \stwo\cdot\hat{\bm L}+
  \stwo\cdot\bmvr\, \sone\cdot\vec{\bm L}\ ,\\
  \srlm&=&\sone\cdot\bmvr\, \stwo\cdot\hat{\bm L}-
   \stwo\cdot\bmvr\, \sone\cdot\vec{\bm L}\ ,
\end{eqnarray}
where $\hat\bmL=\hat\bmr\times\bmp$ is the
``reduced'' orbital angular momentum operator.
In terms of these, $V_{TRV}(\bmr,\bmp)$  can be written as
\bgroup
\arraycolsep=1.0pt
\begin{align}
  V^{({\rm OPE})}(\bmr) &=\frac{\ga g_0^*\mp}{2\fp}\TO\ \ssm \ g'(r)\nonumber\\
  &+
  \frac{g_Ag_1\mp}{4f_\pi}\left(\ttp\ \ssm+\ttm\ \ssp\right)\ g'(r)\nonumber\\
  &+\frac{\ga g_2\mp}{6\fp}\tten\ \ssm \ g'(r)\ ,\\
  V^{({\rm TPE})}(\bmr) &=
  \frac{\ga g_0^*\mp^3}{\fp\Lx^2}\TO\ \ssm\  L'(r)\nonumber\\
  &+\frac{\ga^3 g_0^*\mp^3}{\fp\Lx^2}\TO\ \ssm\  (H'(r)-3L'(r))\nonumber\\
  &-\frac{\ga g_2\mp^3}{3\fp\Lx^2}\tten\ \ssm\  L'(r)\nonumber\\
  &-\frac{\ga^3 g_2\mp^3}{3\fp\Lx^2}\tten\ \ssm\  (H'(r)-3L'(r))\ ,\\
  V^{(3\pi,0)}(\bmr)&=
  -\frac{5\ga^3\Delta_3 M \mp^2}{4\fp\Lx^2}\pi\nonumber\\
  &\times(\ttp\ \ssm+\ttm\ \ssp)(A'(r)+g'(r))\\
  V^{(3\pi,1)}(\bmr)&=\frac{5\ga\Delta_3M\mp^3}{2\fp\Lx^2}(\ttp\ \ssm+\ttm\ \ssp)
  \nonumber\\
  &\Big[4c_1\ L'_\omega(r)-\frac{c_2}{3}(2L'(r)+6L'_\omega(r))\nonumber\\
         &\qquad-\frac{c_3}{2}(3L'(r)+5L'_\omega(r))\Big]\ ,\\
 V^{({\rm CT})}(\bmr) &=  
 {m_\pi^2 \over \Lambda_\chi^2 f_\pi} \Bigl[
   C_1\ \ssm\ Z'(r)+C_2\ \TO\ \ssm\ Z'(r)\nonumber\\
   &+{C_3 \over 2}\ (\ttp\ \ssm-\ttm\ \ssp)\ Z'(r)\nonumber\\
   &+{C_4 \over 2}\ (\ttp\ \ssm+\ttm\ \ssp)\ Z'(r)\nonumber\\
   &+C_5\ \tten\ \ssm\ Z'(r) \Bigr]\ ,
\end{align}
\begin{align}
 V^{({\rm RC})}&(\bmr,\bmp)=\nonumber\\
 &+\frac{\ga g_0^*\mp^3}{2\fp M^2 }\ \TO\ \Big[\frac{\ssm}{4}\Big(g'''(r)
    +2\frac{g''(r)}{r}-2\frac{g'(r)}{r^2}\Big)\nonumber\\
    &+\slm\frac{g'(r)}{r^2}-\frac{g'(r)}{r^2}\ssm L^2
    -\frac{1}{4}\Big(\frac{g''(r)}{r}
    +\frac{g'(r)}{r^2}\Big)\srlm\nonumber\\
    &+\Big(\ssm\Big(g''(r)+2\frac{g'(r)}{r}\Big)
    -i\frac{\srx}{2}\frac{g'(r)}{r}\Big)
    \frac{d}{dr}\nonumber\\
    &+g'(r)\ssm\frac{d^2}{dr^2}\Big]\frac{1}{\mp^2}\nonumber\\
  &+\frac{\ga g_1\mp^3}{4\fp M^2}\
  \ttp\ \Big[\frac{\ssm}{4}\Big(g'''(r)
    +2\frac{g''(r)}{r}-2\frac{g'(r)}{r^2}\Big)\nonumber\\
    &+\slm\frac{g'(r)}{r^2}-\frac{g'(r)}{r^2}\ssm L^2
    -\frac{1}{4}\Big(\frac{g''(r)}{r}
    +\frac{g'(r)}{r^2}\Big)\srlm\nonumber\\
    &+\Big(\ssm\Big(g''(r)+2\frac{g'(r)}{r}\Big)
    -i\frac{\srx}{2}\frac{g'(r)}{r}\Big)
    \frac{d}{dr}\nonumber\\
    &+g'(r)\ssm\frac{d^2}{dr^2}\Big]\frac{1}{\mp^2}\nonumber\\
  &+\frac{\ga g_1\mp^3}{4\fp M^2}\
  \ttm\ \Big[\frac{\ssp}{4}\Big(g'''(r)
    +2\frac{g''(r)}{r}-2\frac{g'(r)}{r^2}\Big)+\nonumber\\
    &+\slp\frac{g'(r)}{r^2}-\frac{g'(r)}{r^2}\ssp L^2
    -\frac{1}{4}\Big(\frac{g''(r)}{r}
    -\frac{g'(r)}{r^2}\Big)\srlp\nonumber\\
    &+\ssp\Big(g''(r)+2\frac{g'(r)}{r}\Big)
    \frac{d}{dr}+g'(r)\ssp\frac{d^2}{dr^2}\Big]\frac{1}{\mp^2}\nonumber\\
   &+\frac{\ga g_2\mp^3}{6\fp M^2}\ \tten\ \Big[\frac{\ssm}{4}\Big(g'''(r)
    +2\frac{g''(r)}{r}-2\frac{g'(r)}{r^2}\Big)\nonumber\\
    &+\slm\frac{g'(r)}{r^2}-\frac{g'(r)}{r^2}\ssm L^2
    -\frac{1}{4}\Big(\frac{g''(r)}{r}
    +\frac{g'(r)}{r^2}\Big)\srlm\nonumber\\
    &+\Big(\ssm\Big(g''(r)+2\frac{g'(r)}{r}\Big)
    -i\frac{\srx}{2}\frac{g'(r)}{r}\Big)
    \frac{d}{dr}\nonumber\\
    &+g'(r)\ssm\frac{d^2}{dr^2}\Big]\frac{1}{\mp^2}\ ,\label{eq:rcoper}
\end{align}
\begin{align}
  &V^{(3\pi-{\rm RC})}(\bmr,\bmp)=
  -\frac{5\ga^3\Delta_3\mp^3}{16\fp\Lx^2}(\ttp\ \ssm+\ttm\ \ssp)\nonumber\\
     &\qquad \Big(\frac{25}{6}L'(r)-\frac{7}{2}L'_\omega(r)+2H'_\omega(r)\Big)
  \nonumber\\
  &\qquad+\frac{25\ga^3\Delta_3\mp^3}{12\fp\Lx^2}\Big[\ttp\ \Big(\Big(\frac{g''(r)}{r}
    +\frac{g'(r)}{r^2}\Big)\srlm\nonumber\\
    &\qquad+2i\srx\frac{g'(r)}{r}\frac{d}{dr}\Big)
    +\ttm\ \Big(\frac{g''(r)}{r}
  -\frac{g'(r)}{r^2}\Big)\srlp\Big]\frac{1}{\mp^2}\ ,\label{eq:rc3pr}
\end{align}
\egroup
with
\begin{eqnarray}
  g(r) &=& \int {{\rm d}^3k\over (2\pi)^3}\; {C_\Lambda(k)\over \mp}
  \frac{1}{\omega_k^2}\,
 e^{i\bmk\cdot\bmr} \ ,\label{eq:g}\\
 L(r) &=& \int {{\rm d}^3k\over (2\pi)^3}\; {C_\Lambda(k)\over m_\pi^3} L(k)\,
 e^{i\bmk\cdot\bmr} \ ,\label{eq:lr}\\
 H(r) &=& \int {{\rm d}^3k\over (2\pi)^3}\; {C_\Lambda(k)\over m_\pi^3} H(k)\,
 e^{i\bmk\cdot\bmr} \ ,\label{eq:hr}\\
 A(r) &=& \int {{\rm d}^3k\over (2\pi)^3}\; {C_\Lambda(k)\over m_\pi^2}
 {s^2\ A(k)\over \omega_k^2}\,
 e^{i\bmk\cdot\bmr} \ ,\label{eq:ar}\\
 L_\omega(r) &=&
 \int {{\rm d}^3k\over (2\pi)^3}\; {C_\Lambda(k)\over m_\pi}
      {L(k) \over \omega_k^2}\,
 e^{i\bmk\cdot\bmr} \ ,\label{eq:lrw}\\
 H_\omega(r) &=&
 \int {{\rm d}^3k\over (2\pi)^3}\; {C_\Lambda(k)\over m_\pi}
      {H(k) \over \omega_k^2}\,
 e^{i\bmk\cdot\bmr} \ ,\label{eq:hrw}\\
 Z(r) &=& \int {{\rm d}^3k\over (2\pi)^3}\; {C_\Lambda(k)\over m_\pi^2}\,
 e^{i\bmk\cdot\bmr} \ .\label{eq:z}
\end{eqnarray}
The functions $g(r)$, $L(r)$, $H(r)$, $A(r)$, $L_\omega(r)$, $H_\omega(r)$
and $Z(r)$
are calculated numerically by
standard quadrature techniques.

\subsection{The $NNN$ potential}\label{sec:NNNr}
In this section we present the explicit derivation
of the $NNN$ potential in configuration space.
Writing explicitly  the integral in Eq.~(\ref{eq:vrnnn}),
neglecting the deltas for simplicity, we get,
\newpage
\begin{align}
  V&(\bmx_1,\bmx_2)=\nonumber\\
  &\quad-\frac{\Delta_3 g_A^3 M}{4\fp^3}\ T_3\
 \frac{1}{2}\int {d^3q\over (2\pi)^3} {d^3Q\over (2\pi)^3}
  e^{-i(\bmq/2)\cdot\bmx_2}\,e^{-i\bmQ\cdot\bmx_1}\,\nonumber\\
  &\quad\times\frac{i(\bmQ+\bmq)\cdot\sone\, i(\bmQ-\bmq)\cdot\stwo\, i\bmQ\cdot\stre}
       {\omega_+^2\omega_-^2\omega_Q^2}\ ,\label{eq:qQxx}
\end{align}
 where $T_3=\tone\cdot\ttwo\ \tau_{3z}+
 \tone\cdot\ttre\ \tau_{2z}+\ttwo\cdot\ttre\ \tau_{1z}$.
 The momenta $\bmq$ and $\bmQ$ in Eq.~(\ref{eq:qQxx}) can be rewritten
 applying the gradient $i\bmna$ to the exponential functions and it reads,
 \begin{align}
   V&(\bmx_1,\bmx_2)=\nonumber\\
   &\frac{\Delta_3 g_A^3 M}{4\fp^3}\ T_3\
     \frac{1}{2}\big[ (\bmna_{\bmx_1}+2\bmna_{\bmx_2})\cdot\sone\
       (\bmna_{\bmx_1}-2\bmna_{\bmx_2})\cdot\stwo\nonumber\\
     &\times\bmna_{\bmx_1}\cdot\stre\big]
     \int{d^3Q\over (2\pi)^3}\ \frac{e^{-i\bmQ\cdot\bmx_1}}{\omega_Q^2}
     \ \int {d^3q\over (2\pi)^3}\ \frac{e^{-i(\bmq/2)\cdot\bmx_2}}
            {\omega_+^2\omega_-^2}\ .\label{eq:qQxxd}
 \end{align}
 The integral in $\bmq$ in Eq.~(\ref{eq:qQxxd}) can be solved using the Feynman tricks.
 The final result is
 \begin{align}
   V(\bmx_1,&\bmx_2)=
   \frac{\Delta_3 g_A^3 M}{4\fp^3}\ T_3\
     \frac{1}{32\pi}\big[ (\bmna_{\bmx_1}+2\bmna_{\bmx_2})\cdot\sone\nonumber\\
       &\times(\bmna_{\bmx_1}-2\bmna_{\bmx_2})\cdot\stwo\
       \bmna_{\bmx_1}\cdot\stre\big]
     \int{d^3Q\over (2\pi)^3}\ \frac{e^{-i\bmQ\cdot\bmx_1}}{\omega_Q^2}\nonumber\\
     &\times \int_{-1}^1 dx\ e^{i(x/2)\bmQ\bmx_2} \frac{e^{-(L/2)x_2}}
           {L}\ ,\label{eq:qxxxd}
 \end{align}
 where $L=\sqrt{Q^2(1-x^2)+4\mp^2}$. The derivative are then evaluated obtaining
 five integral operators $I_i$,
 \begin{eqnarray}
   V(\bmx_1,\bmx_2)&=&\frac{\Delta_3 g_A^3 M}{4\fp^3}\ T_3\ \sum_{i=1,5} I_i
 \end{eqnarray}
 where,
 \begin{align}
   I_1&=-\frac{i}{16\pi}\int{d^3Q\over (2\pi)^3}\
   \frac{e^{-i\bmQ\cdot\bmx_1}}{\omega_Q^2}
   \int_{-1}^1 dx\, \nonumber\\
   &\quad\times e^{i(x/2)\bmQ\cdot\bmx_2}\frac{e^{-(L/2)x_2}}{x_2}
   \,(\sone\cdot\stwo)\bmQ\cdot\stre\ ,\\
   I_2&=+\frac{i}{32\pi}\int{d^3Q\over (2\pi)^3}\
   \frac{e^{-i\bmQ\cdot\bmx_1}}{\omega_Q^2}
   \int_{-1}^1 dx\, (1-x^2) \nonumber\\
   &\quad\times e^{i(x/2)\bmQ\cdot\bmx_2}\frac{e^{-(L/2)x_2}}{L}
   \,\bmQ\cdot\sone\,\bmQ\cdot\stwo\,\bmQ\cdot\stre\ ,\\
   I_3&=-\frac{1}{32\pi}\int{d^3Q\over (2\pi)^3}\
   \frac{e^{-i\bmQ\cdot\bmx_1}}{\omega_Q^2}
   \int_{-1}^1 dx\, (1-x)\nonumber\\
   &\quad\times e^{i(x/2)\bmQ\cdot\bmx_2}e^{-(L/2)x_2}
   \bmQ\cdot\sone\,\bmx_2\cdot\stwo\,\bmQ\cdot\stre\ ,\\
   I_4&=+\frac{1}{32\pi}\int{d^3Q\over (2\pi)^3}\
   \frac{e^{-i\bmQ\cdot\bmx_1}}{\omega_Q^2}
   \int_{-1}^1 dx\, (1+x)\nonumber\\
   &\quad\times  e^{i(x/2)\bmQ\cdot\bmx_2}e^{-(L/2)x_2}
   \,\bmx_2\cdot\sone\,\bmQ\cdot\stwo\,\bmQ\cdot\stre\ ,\\
   I_5&=+\frac{i}{32\pi}\int{d^3Q\over (2\pi)^3}\
   \frac{e^{-i\bmQ\cdot\bmx_1}}{\omega_Q^2}
   \int_{-1}^1 dx\, \Big(L+\frac{2}{x_2}\Big)\nonumber\\
   &\quad\times e^{i(x/2)\bmQ\cdot\bmx_2} \frac{e^{-(L/2)x_2}}{L}
   \,\bmx_2\cdot\sone\,\bmx_2\cdot\stwo\,\bmQ\cdot\stre\ ,
 \end{align}
 To be noticed that integral $I_4=I_3$ exchanging particle 1 and 2.

 In order to divide the angular integration in $\hat{\bmQ}$ from the radial
 integration in $Q$, the exponential functions are expanded in plane waves.
 Therefore the integrals $I_i$ result,
 \begin{equation}
   I_i=\sum_{l,l'}O_i^{ll'}J_i^{ll'}\ ,
 \end{equation}
 where $J_i$ is the integral over $Q$ while in $O_i$
 the integral is over $\hat{Q}$.
 The expressions of $J_i$ are explicitly given below:
 \begin{align}
   J_1^{ll'}&=\frac{1}{8\pi}\int_{-1}^1 dx\ \int_0^\infty dQ \frac{Q^2}{\omega_Q^2}
   \ C_\Lambda(Q)e^{-(L/2)x_2}\nonumber\\
   &\qquad\qquad\times j_l\big({xQx_2\over 2}\big)j_{l'}(Qx_1)\frac{Q}{x_2}
   \ ,\\
   J_2^{ll'}&=\frac{1}{16\pi}\int_{-1}^1 dx\ \int_0^\infty dQ \frac{Q^2}{\omega_Q^2}
   \ C_\Lambda(Q)e^{-(L/2)x_2}\nonumber\\
   &\qquad\qquad\times j_l\big({xQx_2\over 2}\big)j_{l'}(Qx_1)
   \frac{Q^3}{L}(1-x^2)\ ,\\
   J_3^{ll'}&=\frac{1}{16\pi}\int_{-1}^1 dx\ \int_0^\infty dQ \frac{Q^2}{\omega_Q^2}
   \ C_\Lambda(Q)e^{-(L/2)x_2}\nonumber\\
   &\qquad\qquad\times j_l\big({xQx_2\over 2}\big)j_{l'}(Qx_1)
   Q^2(1-x)\ ,\\
   J_4^{ll'}&=\frac{1}{16\pi}\int_{-1}^1 dx\ \int_0^\infty dQ \frac{Q^2}{\omega_Q^2}
   \ C_\Lambda(Q)e^{-(L/2)x_2}\nonumber\\
   &\qquad\qquad\times j_l\big({xQx_2\over 2}\big)j_{l'}(Qx_1)
   Q^2(1+x)\ ,\\
   J_5^{ll'}&=\frac{1}{16\pi}\int_{-1}^1 dx\ \int_0^\infty dQ \frac{Q^2}{\omega_Q^2}
   \ C_\Lambda(Q)e^{-(L/2)x_2}\nonumber\\
   &\qquad\qquad\times j_l\big({xQx_2\over 2}\big)j_{l'}(Qx_1)
   Q\Big(L-\frac{2}{x_2}\Big)\ ,
 \end{align}
 where $j_l(x)$ are spherical Bessel functions and we include a regularization 
 function $C_\Lambda(Q)$ which is defined in Eq.~(\ref{eq:cutoffnnn}).
 The functions $J_i$
 depend only on the modules of $\bmx_1$ and $\bmx_2$ and they 
 are evaluated numerically by standard quadrature techniques.

 For the $O_i$ integrals we get, 
 \begin{align}
   O_1^{ll'}&=(-i)i^{l'}(-i)^{l}\hat{l'}\hat{l}\int d\hat{Q}
   \Big[Y_{l'}(\hat{Q})Y_{l'}(\hat{x}_1)\Big]_0\nonumber\\
   &\times
   \Big[Y_{l }(\hat{Q})Y_{l}(\hat{x}_2)\Big]_0
   \sone\cdot\stwo\, \hat{Q}\cdot\stre\ ,\\
   O_2^{ll'}&=(+i)i^{l'}(-i)^{l}\hat{l'}\hat{l}\int d\hat{Q}
   \Big[Y_{l'}(\hat{Q})Y_{l'}(\hat{x}_1)\Big]_0\nonumber\\
   &\times
   \Big[Y_{l }(\hat{Q})Y_{l}(\hat{x}_2)\Big]_0
   \hat{Q}\cdot\sone\,\hat{Q}\cdot\stwo\,\hat{Q}\cdot\stre\ ,\\
   O_3^{ll'}&=-i^{l'}(-i)^{l}\hat{l'}\hat{l}\int d\hat{Q}
   \Big[Y_{l'}(\hat{Q})Y_{l'}(\hat{x}_1)\Big]_0\nonumber\\
   &\times
   \Big[Y_{l }(\hat{Q})Y_{l}(\hat{x}_2)\Big]_0
   \hat{Q}\cdot\sone\,\hat{x}_2\cdot\stwo\,\hat{Q}\cdot\stre\ ,\\
   O_4^{ll'}&=+i^{l'}(-i)^{l}\hat{l'}\hat{l}\int d\hat{Q}
   \Big[Y_{l'}(\hat{Q})Y_{l'}(\hat{x}_1)\Big]_0\nonumber\\
   &\times
   \Big[Y_{l }(\hat{Q})Y_{l}(\hat{x}_2)\Big]_0
   \hat{x}_2\cdot\sone\,\hat{Q}\cdot\stwo\,\hat{Q}\cdot\stre\ ,\\
   O_5^{ll'}&=(+i)i^{l'}(-i)^{l}\hat{l'}\hat{l}\int d\hat{Q}
   \Big[Y_{l'}(\hat{Q})Y_{l'}(\hat{x}_1)\Big]_0\nonumber\\
   &\times
   \Big[Y_{l }(\hat{Q})Y_{l}(\hat{x}_2)\Big]_0
   \hat{x}_2\cdot\sone\,\hat{x}_2\cdot\stwo\,\bmQ\cdot\stre\ ,
 \end{align}
 which depend only to the angular part of the spatial coordinates
 $\bmx_1$ and $\bmx_2$. The matrix element of the $O_i$ operators
 between HH functions can
 easily expressed in terms of products of 9-j, 6-j and 3-j Wigner
 symbols.

 \section{Details of the calculation and
   convergence of the $a_\Delta(3N)$ coefficient}\label{app:NNNconv}
 In order to compute the even-parity $|\psi^A_{+}\ket$
 and odd-parity $|\psi^A_{-}\ket$ component of the wave function
 we need to solve the following eigenvalue problem,
 \begin{align}\label{eq:egp}
   \begin{pmatrix}
     V_{\rm PC}^{++} & V_{\rm TRV}^{+-}\\
     V_{\rm TRV}^{-+} & V_{\rm PC}^{--}
   \end{pmatrix}
   \begin{pmatrix}
     |\psi^A_{+}\ket\\
     |\psi^A_{-}\ket
   \end{pmatrix}
   =E   \begin{pmatrix}
     |\psi^A_{+}\ket\\
     |\psi^A_{-}\ket
   \end{pmatrix}\,.
   \end{align}
 Using the fact that $||V_{\rm TRV}||<<||V_{\rm PC}||$ we can rewrite
 the eigenvalue problem of Eq.~(\ref{eq:egp}) as,
 \begin{align}\label{eq:egp2}
   \begin{cases}
     V_{\rm PC}^{++}|\psi^A_{+}\ket=E|\psi^A_{+}\ket\\
     |\psi^A_{-}\ket=-(V_{\rm PC}^{--}-E)^{-1} V_{\rm TRV}^{-+}
     |\psi^A_{+}\ket
   \end{cases}\,,
 \end{align}
 where the first equation is the standard eigenvalue problem, while with the
 second we can compute the odd component of the wave function.
 This approach from the numerical point of view results more stable than solving
 directly Eq.~(\ref{eq:egp}) and permits to study the odd component of the
 wave function without solving every time the eigenvalue problem.
  
 Let us study now the convergence pattern of the
 $a_\Delta(3N)$ coefficient, defined in Eq.~(\ref{eq:adelta3}),
 in term of the grandangular  momentum $K$ of the HH basis (for more
 details, see Ref.~\cite{AK08,LE09}). Increasing $K$ is equivalent of
 enlarging the expansion basis. To be definite, in this appendix, we have considered the
 $\tri$  case and used the N4LO/N2LO-500 PC interaction.

 Let us denote with $K^+$  ($K^-$) the maximum value of the
 grandangular momentum of the HH functions used to describe the even
 (odd) part of the wave  function $|\psi^A_{+}\ket$ ($|\psi^A_{-}\ket$). 
 We have computed the even part of the wave function up to complete
 convergence ($K^+=50$) obtaining a binding energy $B(\tri)=8.476$ MeV using the
 first formula in Eq.~(\ref{eq:egp2}).
 For solving the second equation of Eq.~(\ref{eq:egp2}) we have
 performed different calculations varying $K^-$ and we have reported
 the corresponding results for $a_\Delta(3N)$ in Table~\ref{tab:3Nconv}.
 As can be seen by inspecting the table, the pattern of convergence is very
 smooth.  A safe convergence at the third
 digit is reached for $K^-\sim21$, which was the value also
 selected for performing the final calculations. On the other hand, the 
 value of $a_\Delta(3N)$ is not much sensitive to $K^+$, in particular when
 $K^+>20$. Therefore, the value of the coefficient $a_\Delta(3N)$ appears
 to be well under control in our calculation.
 \begin{table}[h]
\begin{center}
\begin{tabular}{cc}
  \hline
  \hline
  $K^-$&  $a_\Delta(3N)$\\
  \hline
   5 &-0.1728 \\
   9 &-0.1842 \\
  13 &-0.1879 \\
  17 &-0.1892 \\
  21 &-0.1897 \\
 \hline
 \hline
\end{tabular}
\caption{ \label{tab:3Nconv}
  Convergence pattern of the $\tri$  $a_\Delta(3N)$ coefficient as function of $K^-$,
  the maximum grandangular momentum of the HH
  functions used for constructing the odd-parity component
  of the $\tri$ wave function. For the even-parity we
  have used $K^+=50$, a value sufficient to reach full convergence
  of the even-parity component of the wave function.
  The reported calculations are performed
  using N4LO/N2LO-500 PC interaction.}
\end{center}
\end{table}

\end{document}